\documentclass[preprint]{aastex}
\shorttitle{Correlation analysis of NLS1 galaxies}
\shortauthors{Xu et al.}

\newcommand{\kms}{km s$^{-1}$}
\newcommand{\oiii}{[\ion{O}{3}]}

\newcommand{\sii}{[\ion{S}{2}]}

\newcommand{\feii}{\ion{Fe}{2}}

\newcommand{\hb}{H$\beta$}
\newcommand{\hbb}{H$\beta_{\rm b}$}

\newcommand{\LLedd}{$L/L_{\rm Edd}$}
\newcommand{\Mbh}{$M_{\rm BH}$}

\newcommand{\n}{$n_{\rm e}$}

\newcommand{\cm}{cm$^{-3}$}
\newcommand{\oiiicore}{[\ion{O}{3}]$_{\rm c}$}
\newcommand{\voiii}{$v_{\rm [OIII]_c}$}

\begin{document}

\title{Correlation analysis of a large sample of narrow-line Seyfert 1 
galaxies: linking central engine and host properties}

\author{Dawei Xu}
\affil{National Astronomical Observatories, Chinese Academy of
       Sciences, 20A Datun Road, Beijing 100012, China; dwxu@nao.cas.cn}

\author{S. Komossa}
\affil{
National Astronomical Observatories, Chinese Academy of
Sciences, 20A Datun Road, Beijing 100012, China;
Technische Universitaet Muenchen, Lehrstuhl fuer Physik I,
James Franck Strasse 1/I, 85748 Garching, Germany;
Excellence Cluster Universe, TUM, Boltzmannstrasse 2, 85748
Garching, Germany
}

\author{Hongyan Zhou} 
\affil{ 
Key Laboratory for Research in Galaxies and Cosmology, 
University of Sciences and Technology of China, 
Chinese Academy of Sciences, 96 JinZhai Road, Hefei 230026, China;
Center for Astrophysics, University of Science and 
Technology of China, 96 JinZhai Road, Hefei 230026, China;
Polar Research Institute of China, 
451 Jinqiao Road, Pudong, Shanghai 200136, China
}

\author{Honglin Lu}
\affil{
Key Laboratory for Research in Galaxies and Cosmology, 
University of Sciences and Technology of China, 
Chinese Academy of Sciences, 96 JinZhai Road, Hefei 230026, China;
Physics Experiment Teaching Center, 
University of Sciences and Technology of China, 96 JinZhai Road, 
Hefei 230026, China
}

\author{Cheng Li}
\affil{
Partner Group of the MPI f\"ur Astrophysik at Shanghai Astronomical 
Observatory, 
Key Laboratory for Research in Galaxies
and Cosmology of Chinese Academy of Sciences, 
80 Nandan Road, Shanghai 200030, China
}

\author{Dirk Grupe}
\affil{Department of Astronomy and Astrophysics, Pennsylvania State University, 525 Davey Lab, University Park, PA 16802, USA}

\author{Jing Wang}
\affil{National Astronomical Observatories, Chinese Academy of
       Sciences, 20A Datun Road, Beijing 100012, China}
\and
\author{Weimin Yuan}
\affil{National Astronomical Observatories, Chinese Academy of
       Sciences, 20A Datun Road, Beijing 100012, China}

\begin{abstract}
We present a statistical study of a large, homogeneously analyzed sample
of narrow-line Seyfert 1 (NLS1) galaxies, accompanied
by a comparison sample of broad-line Seyfert 1 (BLS1) galaxies.
Optical emission-line and continuum properties are subjected
to correlation analyses, in order to identify the main drivers of
active galactic nuclei (AGN) correlation space, and of NLS1 galaxies
in particular. For the first time, we have established 
the density of the narrow-line region
as a key parameter in Eigenvector 1 space, as important as 
the Eddington ratio \LLedd.
This is important because it 
links the properties of the central engine
with the properties of the host galaxy; i.e., the interstellar medium (ISM).
We also confirm previously found correlations involving
the line width of \hb,
and the strength of the \feii\ and \oiii\ $\lambda$5007
emission lines,
and we confirm the important role played by 
\LLedd\ in driving the properties of NLS1 galaxies. 
A spatial correlation analysis shows that large-scale
environments of the BLS1 and NLS1 galaxies of our sample are similar.
If mergers are rare in our sample, accretion-driven winds on the one
hand, or bar-driven inflows on the other hand, may account
for the strong dependence of Eigenvector 1 on ISM density.
\end{abstract}


\section{Introduction}

Narrow-line Seyfert 1 (NLS1) galaxies as active galactic nuclei (AGN) 
with the narrowest Balmer lines from
the broad-line region (BLR) and the strongest
\feii\ emission, cluster at one extreme end of
AGN correlation space.
It is expected that such correlations provide some
of the strongest constraints on, and new insights in, the physical conditions
in the centers of AGN and the prime drivers of activity,
and the study of NLS1 galaxies is therefore of
particular interest (see Komossa 2008 for a review).        

A number of key results have arisen from principal component analyses (PCA).
Applied to NLS1 and broad-line Seyfert 1 (BLS1) galaxies,
the strongest correlations are among optical AGN properties,
and involve the width of
the H$\beta$ line and the strength of the \oiii\ line and \feii\ complex.
Others include the steepness of the X-ray spectrum and the asymmetry
of the CIV emission line
(e.g. Boroson \& Green 1992, Laor et al. 1997, Brandt 1999,
Marziani et al. 2001, Sulentic et al. 2000, 2002, Boroson 2002,
Grupe 2004, Sulentic et al. 2007, 2008).

These multi-wavelength observations (and their interpretation or modeling)
hint at smaller black hole masses
in NLS1 galaxies, and as such their black holes represent
an important link with the elusive intermediate mass black holes, which
have been little studied so far. Accreting likely
at very close to the maximum allowed values, NLS1 galaxies are
important test-beds of accretion models.  

Given their low black hole masses, mergers of NLS1 galaxies are prime sources
of gravitational wave emission within the LISA sensitivity band
(e.g., Centrella 2010).

Independent samples are of importance when assessing the robustness
of correlation analyses, and to increase correlation space (Grupe et al.
1999, V\'{e}ron-Cetty et al. 2001, Williams et al. 2002, Zhou et al. 2006,
Grupe et al. 2010). 
We have studied a sample of 94 AGN, including 55 NLS1 galaxies and 39 
BLS1 galaxies. Ours is a large, homogeneously analyzed
sample, and we add new emission-line measurements (particularly, 
the density-sensitive \sii\ ratio) to correlation analyses. 
The comparison sample of BLS1 galaxies enables us to study trends across
the whole AGN population; and to distinguish between physical processes
operating only within the NLS1 population, and those which are 
relevant for AGN as a class. 

This paper is the fourth in a sequence. The first one
concentrated on the differences in the narrow-line region (NLR) density
of BLS1 and NLS1 galaxies (Xu et al. 2007, hereafter X07), the second
demonstrated that NLS1 and BLS1 galaxies follow
the same $M_{\rm BH}-\sigma$ relation if ``blue outliers'' 
are removed from the sample (Komossa \& Xu 2007, hereafter KX07),
and the third discussed the properties of Seyfert galaxies with 
extreme outflows (``blue outliers''; Komossa et al. 2008, hereafter K08). 
In this fourth paper of the sequence,
we focus on sample properties and correlation analyses. 
This paper is organized as follows. In Section 2 we describe the 
sample selection and the methods
of data analysis.  Section 3 reports the results obtained
from the emission-line and continuum measurements. 
In Section 4 and 5 results from our
correlation analyses and cross-correlation analysis are given, 
which are then discussed and interpreted in Section 6.
We conclude with a summary in Section 7.  
 
We use the terms NLS1 galaxies and BLS1 galaxies collectively for
high-luminosity and low-luminosity objects, i.e., 
jointly for Seyfert galaxies and quasars.
Throughout this paper, a cosmology with $H_{\rm 0}=70$\,\kms\,Mpc$^{-1}$,
$\Omega_{\rm M}=0.3$ and $\Omega_{\rm \Lambda}=0.7$ is adopted.

\section{Data analysis}

\subsection{The sample}
Our sample consists of NLS1 galaxies 
from the catalog of V\'{e}ron-Cetty \& V\'{e}ron (2003),
to which we added  
a comparison sample of BLS1 galaxies 
from Boroson (2003) 
at $z < 0.3$, first presented by X07.
All galaxies have been observed in the
course of the {\it Sloan Digital Sky Survey} (SDSS) 
(Data Release 3; Abazajian et al. 2005)
and have detectable low-ionization emission lines (in particular,  
\sii\ $\lambda\lambda$6716, 6731 is always present with S/N $>$ 5).
The sample of NLS1 galaxies has similar redshift and absolute 
magnitude distribution as the BLS1 galaxy sample.
The original sample selection, data preparation, 
and data analysis methods are described in detail 
by X07. Here, we briefly summarize the procedures
in Sect. 2.2 and 2.3. For the first time, we present our
measurements for the full sample in table form (Table\,1).
While the focus of our study is on NLS1 galaxies, having a comparison
sample of BLS1 galaxies is important. Our BLS1 sample contains
approximately equal numbers as the NLS1 sample. While this is
the best approach for comparison purposes, we also note that
in observations, NLS1 galaxies make only approximately 20\% of
the Seyfert 1 population.

\subsection{Decomposition of starlight and nuclear continuum} 
All SDSS spectra were corrected 
for Galactic extinction, the continuum was decomposed
into host galaxy 
and AGN components (Lu et al. 2006), and then we subtracted
the starlight component,  
nuclear continuum and the \feii\ complexes from the spectra.
More precisely, 
the spectra were decomposed into the following four components 
(see X07 for details):
(1) A starlight component modeled by 6 synthesized galaxy
    templates, which were built from the synthetic spectral library of Bruzual
    \& Charlot (2003). 
The details of the algorithm were described in Lu et al. (2006).
(2) A power-law continuum to represent the AGN continuum. 
(3) A Balmer continuum generated in the same way as Dietrich et al. (2003).
(4) An \feii\ template given by V\'{e}ron-Cetty \& V\'{e}ron (2004).
The final multicomponent fit including the \feii\ complexes 
was then subtracted from 
the observed spectrum and the emission line properties
were measured in the way described in Sect.\,2.3.
 
\subsection{Emission line fits}
Emission line profiles of the galaxies were fit 
with Gaussians and/or Lorentzians using the IRAF package 
SPECFIT (Kriss 1994).
Measured FWHMs were corrected for instrumental broadening.
Complex emission lines, which cannot be represented by
a single Gaussian profile, were decomposed into multiple components.   
The Balmer lines were decomposed into a narrow and a broad
component representing emission from the NLR and 
BLR, respectively.
The narrow core was fit employing a single Gaussian profile with
FWHM fixed to that determined for \sii\ $\lambda\lambda$6716, 6731. 
As in previous work (e.g., Rodriguez-Ardila et al. 2000, 
Dietrich et al. 2005, X07, Mullaney \& Ward 2008),
the broad component was fit by using a combination of 
two Gaussian profiles, or alternatively a single Lorentzian profile
(e.g., V\'{e}ron-Cetty et al. 2001, Sulentic et al. 2002).
For the approach employing the two-Gaussian profile fit, the final width of the 
broad-line emission, \hbb, is determined as the FWHM of the sum of
the two Gaussians. We do not assign a physical meaning to
the two separate broad components; 
they merely serve as a mathematical description of 
the complex shapes of the broad lines.
While comparable and equally reasonable results can be achieved with both
single Lorentzian and two Gaussian profiles
for most broad components of the Balmer lines of NLS1 galaxies,
generally no acceptable fit is possible when
employing the Lorentzian profiles to fit BLS1 galaxies.
Among the NLS1 galaxies, 
line width measurements between Lorentzian and
Gaussian fits typically agree within 20\%, Lorentzian profiles
always resulting in smaller FWHMs, as expected. 
We finally used the results of the multi-{\em Gaussian} fits 
(Table\,1) for the NLS1
classification and also for further correlation analysis.
This allows a direct comparison with the BLS1 control sample, and also
with several previous studies (e.g. Boroson \& Green 1992, 
Grupe et al. 1999, Vaughan et al. 2001).

Apart from the Balmer lines, \oiii\ shows a complex profile. 
The total \oiii\ profile, \oiii$_{\rm total}$ was decomposed into two Gaussian 
components: a narrow core and a broad base. 
The narrow core of \oiii\ is referred to as \oiiicore. 
Measurements of the FWHM and blueshift of \oiii\ (Table\,1)
refer to the core of \oiii, unless noted otherwise.
The velocity shift of \oiii\ was measured relative to \sii. 
We use positive velocity values to refer to blueshifts, negative ones for
redshifts.
All other forbidden lines were well fit with single Gaussian profiles.  

The strength of the \feii\ emission, R4570, was measured as the ratio
of the flux of the \feii\ complex between the rest wavelengths 
4434 and 4684 \AA\ to that of total \hb\ emission.
The strength of the \oiii\ emission, R5007, was measured as  
the flux ratio of total \oiii\ $\lambda$5007 over total \hb\ emission.
``Total'' \hb\ emission refers to the sum of broad 
and narrow component.  

\subsection{A note on NLS1 classification}
After re-classification based on spectral emission-line fitting, we have 
39 BLS1 and 55 NLS1 galaxies in our sample.
The standard classification criterion of NLS1 galaxies
is according to FWHM(\hbb) $<$ 2000 \kms\ (Goodrich 1989)
which is most commonly applied.  
Such NLS1 galaxies usually come with weak 
\oiii$\lambda5007$/\hb\ and strong \feii/\hb\
which form part of the classification criteria 
(Osterbrock \& Pogge 1985, Goodrich 1989, V\'{e}ron-Cetty et al. 2001).
We note in passing that Sulentic et al. (2001) rather suggest to
use FWHM(\hbb) $=$ 4000 \kms\  as `dividing' value,
because they find AGN properties to change more significantly at that value. 
Furthermore, a few authors pointed out that NLS1 classification should 
incorporate a luminosity dependence of the FWHM(\hbb) cut-off
(e.g., Laor 2000, V\'{e}ron-Cetty et al. 2001,
Shemmer et al. 2004).
Netzer \& Trakhtenbrot (2007) suggest to base NLS1 classification
on the Eddington ratio, and use $L/L_{\rm Edd} \ge 0.25$ for NLS1 galaxies.  
In the current study, we continue to use the classical FWHM cut-off of 
FWHM(\hbb) $<$ 2000 \kms\ which is still
most commonly used. Results (Table\,1) are reported 
in dependence of FWHM(\hbb) and $L/L_{\rm Edd}$, 
so that other cut-off values could be easily applied to our sample. 

\section{Emission-line and AGN properties}

Results from our emission-line and continuum measurements are listed 
in Table\,1. 
Relevant parameters of the continua and emission lines were used to derive 
some AGN parameters, including black hole masses and Eddington ratios
(Figure\,1). 
We describe below how we determined these parameters.

Assuming that the motion of the BLR clouds is virialized (e.g., 
Wandel et al. 1999), the black hole mass can be estimated as 
$M_{\rm BH} = R_{\rm BLR}v^2/G$. The velocity $v$ of the 
BLR clouds can be estimated from the FWHM 
as $v=\sqrt{3}/2 \times {\rm FWHM}$,
by assuming an isotropic cloud distribution. 
There is only few reverberation mapping results for NLS1 galaxies 
so far (Peterson et. al. 2000, Bentz et al. 2009).
We assumed that the relation between the radius of the BLR and
the optical luminosity, established from reverberation mapping for 
nearby Seyfert galaxies (e.g., Peterson et al. 2004; 
Kaspi et al. 2005) is indeed applicable to NLS1 galaxies 
(e.g., Komossa et al. 2006). 

We estimated the black hole masses of our NLS1 and BLS1 galaxies using 
the $R_{\rm BLR}$--$\lambda L_{5100}$ relation 
given by Kaspi et al. (2005). 
The luminosity $L_{5100}$ was derived from the SDSS photometry%
\footnote{
The SDSS photometry calibration is more robust given that the imaging 
data is taken on moonless photometric nights under good seeing 
conditions (Hogg et al. 2001), while spectroscopy is
done on those nights that are not photometric, or with seeing worse
than 1.7$\arcsec$ FWHM, or with a moderate amount of moon 
(e.g., Abazajian et al. 2003). 
Moreover, the SDSS spectrophotometric fluxes before Data Release 6 were tied 
to fiber magnitudes that led to an under-estimation in the spectrophotometric 
scale (Adelman-McCarthy et al. 2008). 
Magnitudes used to derive black hole masses 
include a contribution from host galaxies.
However, most of our objects are AGN dominated and thus the contamination
from host galaxies is small.  
We compared our fluxes derived from the SDSS photometry to the 
spectroscopic fluxes and found that differences of flux scale between 
the two calibrations can reach at maximum 35\%.
}. 
The PSF $g$ and $r$ magnitudes (Galactic extinction corrected)
were used to determine the continuum slope (e.g., Wu \& Liu 2004) 
and the flux density at 5100\,\AA\ rest frame.
The FWHM of the broad component from the multi-{\em Gaussian}
fit of \hb, FWHM(\hbb) (Table\,1), was used for
the black hole mass estimates reported in Table\,1. 
An error in black hole mass is typically 0.5 dex,
which arises from the use of single-epoch data
(e.g., Vestergaard 2004),
and uncertainties in H$\beta$ decomposition
and in host galaxy contribution.
 
The NLS1 galaxies of our sample have, on average, smaller black hole masses
than BLS1 galaxies. 
The estimated black hole masses of NLS1 galaxies range from
$\log M_{\rm BH, NLS1} = 5.7\,{\rm M}_{\odot}$ to $7.3\,{\rm M}_{\odot}$
with an average value of $6.5\,{\rm M}_{\odot}$, while 
BLS1 galaxies range from $\log M_{\rm BH, BLS1} = 6.5\,{\rm M}_{\odot}$
to $8.4\,{\rm M}_{\odot}$ with an average value of $7.2\,{\rm M}_{\odot}$
(Figure\,1).

Eddington ratios, represented by \LLedd, 
were calculated from the deduced black hole masses, according to 
$L_{\rm Edd} = 1.3\,10^{38}$\,$M_{\rm BH}/M_{\odot}\, {\rm erg\,s}^{-1}$.
For bolometric luminosity correction, we assumed 
$L_{\rm bol}=9\,\lambda{L_{\rm 5100 }}$ (Kaspi et al. 2000).
As shown previously, NLS1 galaxies exhibit, on average, 
higher Eddington ratios than BLS1 galaxies.
Eddington ratios of the NLS1 galaxies of our sample range from
$\log$ \LLedd$_{\rm, NLS1} = -0.6$ to $0.3$ with an average value of $-0.1$,
while BLS1 galaxies range from $\log$ \LLedd$_{\rm, BLS1} = -1.4$ to $-0.3$
with an average value of $-0.8$.

The distributions of black hole masses and Eddington ratios 
are shown in Figure\,1a and 1b, respectively.
We have also applied a Kolmogorov-Smirnov (K-S) test, in order to 
confirm that the two distributions
for NLS1 and BLS1 galaxies are significantly different. Indeed, 
for both the distributions, we find a K-S probability of ${P < 10^{-4}}$ 
that the two samples are drawn from the same parent population.
For comparison, we also plot histograms of other properties (Figure\,2);
some of these reflecting trends we reported previously.
(1) The distribution of FWHM(\sii) (Figure\,2a) shows
narrower \sii\ in NLS1 galaxies than BLS1 galaxies (KX07).
(2) The distribution of \sii\ intensity ratios and the NLR density 
(Figure\,2b and 2c) demonstrate our discovery of 
a zone of avoidance in density, in the sense that BLS1 galaxies avoid
low densities, while NLS1 galaxies show a larger scatter in density,
including a significant number of objects with low densities (X07).
(3) The distribution of FWHM(\oiiicore) reflects that
NLS1 galaxies have similar \oiiicore\ widths as BLS1
galaxies (Figure\,2d). 
However, if we exclude galaxies with high velocity shifts 
(i.e., \voiii\ $>$ 75 \kms), NLS1 and
BLS1 galaxies show different scatter 
(the inset of Figure\,2d)
(KX07).
(4) The distribution of velocity shifts of \oiii\ core lines 
(Figure\,2e) reflects that NLS1 and BLS1 galaxies
differ in \oiii\ outflow velocity, in the sense
that NLS1 galaxies have higher blueshifts of \oiii\ core lines than
BLS1 galaxies, including a number of ``blue outliers'' with 
extreme blueshifts (K08). 
(5) The distributions of optical \feii\ and \oiii\ strength (Figure\,2f and
2g) indicate that NLS1 galaxies have stronger \feii\ emission than 
BLS1 galaxies, and similar \oiii\ emission as BLS1 galaxies  
(X07).

We list in Table\,1 the key properties of our sample.
Column (1) gives celestial coordinates in RA and DEC.
Column (2) gives galaxy name.
Column (3) gives the redshift. 
Column (4) lists the SDSS $i$-band absolute magnitude, derived from the 
PSF $i$ magnitude.
Column (5) lists the FWHM of the broad component of H$\beta$ in \kms.
Column (6) lists the ratio of total \oiii\ over total H$\beta$ emission.
Column (7) lists the ratio of \feii\ 4570 over total H$\beta$ emission.
Column (8) lists the FWHM of \sii\ in \kms. 
Column (9) lists the FWHM of the core of \oiii\ in \kms.
Column (10) lists the \oiiicore\ velocity shift (blueshift) with 
respect to \sii\ in \kms.
Column (11) lists the \sii\ intensity ratio of \sii\ $\lambda6716/\lambda6716$.
Column (12) lists the NLR electron density in cm$^{-3}$,
estimated from the density-sensitive \sii\ intensity ratio 
(see X07 for details).
Column (13) lists the flux density at 5100\,\AA\, restframe in
10$^{-17}$~erg~s$^{-1}$~cm$^{-2}$~\AA$^{-1}$.
Column (14) lists the $\log$ of the monochromatic luminosity at 5100\,\AA.
Column (15) lists the $\log$ of the Eddington ratio. 
Column (16) lists the $\log$ of the black hole mass.
The key sample properties of NLS1 and BLS1 galaxies are summarized in Table\,2.

\section{Correlation analyses}

We now focus on studying trends across our sample,
in order to perform a comparison with previously known correlations
derived for different NLS1 samples, and to identify new trends.
We have performed two types of correlation analyses.
First, a Spearman rank order correlation analysis was used to
determine which parameters are correlated and
to derive the significance of the correlations.
Second, we have performed a principle component analysis
(PCA; e.g. Pearson 1901, Duda et al. 2001).
The PCA is a commonly applied method aimed at
identifying the strongest correlations and the underlying parameters.
This method allows us to
link together diverse correlations into meaningful groups
and reveal the physical mechanisms driving these correlations
(for further discussion, and some caveats, see Boroson 2004).

\subsection{Spearman rank order correlations}

In a first step we computed Spearman rank order correlation coefficients
between 11 of the key properties of our sample
which were given in Table\,1.
These correlations involve emission-line widths
(FWHM(\hbb), FWHM(\sii) and  FWHM(\oiiicore)),
emission-line ratios (R5007, R4570, the ratio R${\rm_{[SII]}}$
defined as the intensity ratio of \sii\ $\lambda 6716/\lambda 6731$, 
and the inferred NLR density \n),
and the parameters $\lambda{L_{\rm 5100}}$, \LLedd, and \Mbh.

We note in passing that we have included both parameters,
R${\rm_{[SII]}}$ and \n, in the correlation analysis. Even though
strongly dependent on each other, their dependence is
not linear, and we have therefore kept both parameters.
Correlations with \n\ are slightly stronger than with R${\rm_{[SII]}}$.

Table\,3 shows the results of our correlation analysis.
We have split our sample into three groups; the NLS1
galaxies (55 objects), the BLS1 galaxies (39 objects), and the whole sample,
and we have run the correlation analysis separately for
each of the three, and separately report correlation coefficients
in Table\,3.
All correlations with Spearman rank probabilities $P_{\rm s} < 0.01$
are marked in boldface. 
Figure\,3 displays the correlation diagrams for several key parameters.

The strongest (anti-)correlation among independent
parameters involves the
strength of \feii\ over total \hb\ emission, R4570.
Among directly observed parameters,
the clearest anti-correlation is between R4570
and the width of the broad component of \hb, FWHM(\hbb).
The trend is significant for the whole sample, with a
Spearman rank order correlation coefficient of
$r_{\rm s}=-0.7$.
This anti-correlation remains strong among the NLS1 and BLS1 galaxy
sub-populations.
In addition,  we also confirm the
known anti-correlation
between the strengths of \feii\ (R4570) and \oiii\ (R5007), 
which again is apparent in the
full sample, and the two subsamples.
The third strong correlation involving R4570, which persists
across the BLS1 and NLS1 subsample, is the one between
R4570 and \LLedd\  ($r_{\rm s}=0.7$). That correlation is
stronger than with luminosity (no correlation at all)
or with black hole mass, suggesting that \LLedd\ is the
more fundamental parameter.

Two further correlations which involve the width of \hbb\ 
echo the trend we have already reported previously:
the correlation between FWHM(\hbb) and density, which becomes
apparent for the full sample (X07), and the correlation
between \voiii\ and FWHM(\oiiicore)
(K08), which is prominent in NLS1 galaxies ($r_{\rm s}=0.6$),
but absent in BLS1 galaxies.

We have also investigated correlations among
the widths of \hbb, \sii, and \oiii.
While there is no correlation between FWHM(\hbb) and
FWHM(\sii) for each subsample, such a trend
becomes apparent across the full sample.
The finding that a
correlation between FWHM(\sii) and FWHM(\oiiicore) 
is significant
among BLS1 galaxies with $r_{\rm s}=0.67$,  but does not exist at
all among the NLS1 galaxies, was already explained
by K08: a subsample
of NLS1 galaxies (so called blue-outliers) have a peculiarly broadened
extra-component in \oiii\ (K08), and these few blue outliers destroy the
expected correlation.
If we remove from our NLS1 sample all those galaxies which
show evidence for blue outliers (explicitly, objects with
\oiii\ blueshifts exceeding \voiii\ $> 75$ \kms),
then we also find a correlation between FWHM(\sii) and 
FWHM(\oiiicore) in our NLS1 sample ($r_{\rm s}=0.47$).

We note that the width of \hbb\ does not systematically
increase with continuum luminosity $\lambda L_{\rm 5100}$. 
This shows that NLS1 classification does not strongly depend on 
luminosity (Figure 4a). We also note in passing, that all
of our NLS1 galaxies fulfill the criterion of Netzer \& Trakhtenbrot 
(2007) of $L/L_{\rm Edd} \ge 0.25$ (Figure\,4b).

We would like to recall here the well-known fact that
correlation analyses depend on the parameters considered, and they
may not reveal the ultimate underlying correlation or driver. 
Two parameters
may simply correlate because both depend on a third parameter which
was not considered in the analysis.  However, simple correlation analysis
is a useful first step in searching for trends across data sets,
and is a commonly used method, and in that sense we also applied it.
As next step toward the final goal of identifying the
strongest correlations in our sample and finding the key physical
drivers behind them, we have applied a principal component analysis.

\subsection{Principal component analysis (PCA)}

PCA is a mathematical technique to
convert a set of observations of possibly correlated variables
into a set of values of uncorrelated variables called principal components,
or eigenvectors.
The eigenvectors as a whole reproduce the original data space, and are
orthogonal to each other.
Eigenvector 1 (hereafter EV1) accounts for the largest
variance of the data among the original measured variables, and
Eigenvector 2 (hereafter EV2), being orthogonal to (uncorrelated with)
the first, represents the second greatest variance.
Eigenvalues are defined as the fraction of the variance that
each eigenvector accounts for.
A good account of the application of PCA in astronomy
can be found in Francis \& Wills (1999).

In a second step, we have run a PCA%
\footnote{we used the software available at
http://www.classification-society.org/csna/mda-sw/}
on the following input parameters:
$M_{\rm i}$, FWHM(\hbb), R5007, R4570, 
FWHM(\sii), $v_{\rm_{[OIII]_c}}$ and R${\rm_{[SII]}}$.
Each parameter represents independent information.
Note that for the first time we involve the NLR density (reflected
by the ratio R${\rm_{[SII]}}$) in the PCA.
The results of the PCA are summarized in Table\,4
(see also Figure\,5 and Figure\,6).

We find that EV1 accounts for 38\% of
the variance in the data.
EV1 is significantly (anti-)correlated with R4570, FWHM(\hbb),
R${\rm_{[SII]}}$, FWHM(\sii) and $v_{\rm_{[OIII]_c}}$.
When EV1 decreases, R4570 increases, outflow
velocity becomes stronger and R${\rm_{[SII]}}$ increases,
while R5007, FWHM(\hbb) and FWHM(\sii) decrease.
EV2 accounts for 18\%
of the variance. EV2 is mostly anti-correlated with
the absolute $i$ magnitude and correlated with FWHM(\sii).

In order to understand the relationship between the physical properties
of AGN and the first two prime eigenvectors,
we test correlations between EV1 and EV2
and \n, $\lambda{L_{\rm 5100}}$, \Mbh, and \LLedd, respectively
(Figure\,5),
using the Spearman rank correlation test (Table\,5).
It can be seen that EV1 of our sample strongly correlates with
\n, in addition to \LLedd\ and \Mbh, while EV2 with 
$\lambda{L_{\rm 5100}}$ and \Mbh.

Figure\,6 shows the distribution of the objects 
of our sample in the EV1-EV2 space.
In order to examine the dependence of the distribution on
different parameters, each parameter is 
divided into three bins and coded by symbol size.
The size of each point refers to coding according to 
\oiii$_{\rm c}$ blueshift, R$_{\rm [SII]}$, FWHM$_{\rm [SII]}$,
R4570, $L/L_{\rm Edd}$, $M_{\rm BH}$, respectively. 
It is interesting to note that some of the objects with extreme
values of EV1 and EV2 can be isolated from the EV1-EV2 diagram.
We find that several ``blue outliers'' are located at 
the upper left corner of the figure (Figure\,6a).
This regime is also dominated by objects with strong 
\feii\ emission and high $L/L_{\rm Edd}$ (Figure\,6d and 6e). 
We also note that objects with high R$_{\rm [SII]}$ fall at 
the low end of EV1 of the diagram (Figure\,6b),
and objects with high black hole mass are found preferentially at 
the high end of EV1 (Figure\,6f).

\section{The large-scale environment}

In order to probe the large-scale environment of the galaxies
in our sample, and to search for possible differences in the
environment of the NLS1 galaxies in comparison to the BLS1 galaxies,
we have used the projected
redshift-space two-point cross-correlation function (2PCCF), $w_p(r_p)$.
We calculated the 2PCCF between the NLS1 or BLS1
galaxy samples and a reference sample of about half a million
galaxies selected from the main spectroscopic sample of the 
SDSS final data
release (DR7; Abazajian et al. 2009). A random sample was 
constructed so as to have the same selection effects as the 
reference sample. The reference and random samples were 
cross-correlated with the same set of galaxies (the NLS1 or BLS1
galaxy samples), and $w_p(r_p)$ as a function of the projected
separation $r_p$ was defined by the ratio of the two pair
counts minus one. Details about our methodology for computing
the correlation functions and for constructing the reference
and random samples can be found in Li et al. (2006). For
consistency, we restricted our reference galaxies to the same
redshift range as our NLS1/BLS1 samples, i.e., $z<0.3$

The amplitude of 2PCCF on scales larger than a few Mpc provides
a direct measure of the mass of the dark matter halos that 
host the galaxies through the halo mass-bias relation. As shown
in Li et al. (2008), the amplitude of the correlation function
on scales $\lesssim 100$ kpc can serve as a probe of physical
processes such as mergers or galaxy-galaxy interactions. On 
intermediate scales, the correlation probes the so-called
'1-halo' term where the pair counts are dominated by galaxy 
pairs in the same dark matter halo. Therefore, the 2PPCF is
a powerful measure of ``environment'' in the sense that it 
encapsulates information about how galaxy properties depend
on the overdensity of the galaxy environment over a wide range
of physical scales.

In Figure\,7 we compare the 2PCCFs estimated
for our NLS1 and BLS1 galaxies. We do not see significant
difference in $w_p(r_p)$ between the two classes of galaxies,
over all the spatial scales probed ($10kpc/h<r_p<30Mpc/h$).
This indicates that, to the first order, NLS1 and BLS1 galaxies are
found in similar environments.%
\footnote
{It is well known that the clustering of galaxies
strongly depends on their host properties. 
In order to perform a more detailed comparison,
than we have done so far, a close matching of all
involved samples in host galaxy properties, like 
luminosity, color and morphology, is required.} 

\section{Discussion}

Previous observations and their interpretation strongly indicate
that NLS1 galaxies are AGN with low mass black holes typically accreting
near the Eddington rate. As such, NLS1 galaxies as a class may hold
important clues on black hole growth and evolution (e.g., Mathur et al. 2001),
and on feeding and feedback in the course of galaxy evolution.
We have therefore studied and homogeneously analyzed a large
sample of NLS1 galaxies, accompanied by a comparison sample of
BLS1 galaxies analyzed in the same way.
This paper of the sequence has focused on correlation analyses,
and PCA in particular.

\subsection{PCA and Eigenvector space} 

PCA has previously turned
out to be a powerful tool in uncovering trends
and correlations among AGN samples
(e.g., Boroson \& Green 1992, Brandt 1999, Sulentic et al. 2000, 2002,
Marziani et al. 2001, Grupe 2004, Wang et al. 2006, Sulentic et al. 2007,
Zamfir et al. 2008, Mao et al. 2009,
Kuraszkiewicz et al. 2009,
Marziani et al. 2010; see Boroson 2004 for
a critical assessment of advantages and shortcomings of this method).
However, it has been also noted that PCA
are specific to the samples examined, depending on the
observed properties used,
as well as on the ranges of the parameters (e.g., Grupe 2004).
Different eigenvectors may emerge from different samples.
Adding new elements to PCA is therefore of great importance
to verify the reality of the correlations, and furthermore
to unveil the physical properties which determine the observable
characteristics of AGN.

The most popular interpretation of EV1 by far has been
that EV1 is likely to be driven by the Eddington ratio \LLedd\
(e.g., Sulentic et al. 2000, Boroson 2002, Grupe 2004).
For the first time, we included the NLR density in the PCA, and have 
shown that EV1 of our sample is highly correlated
with density, in addition to \LLedd.
EV2 of our sample is dominated by luminosity and
therefore accretion rate (proportional to luminosity),
which matches the interpretation of EV2 of Boroson (2002).

One key advantage of PCA is to discriminate between the
various classes of objects according to their loci
on the EV1-EV2 plane.
The identification of objects that represent extremes
are particularly crucial because their properties
may suggest the nature of the physical parameter that
governs the correlations (e.g., Boroson 2004).
The similarities of the first two eigenvectors of
our sample and Boroson (2002) allow a comparison of
extreme objects in the EV1-EV2 diagram.

We earlier found that NLS1 galaxies have higher blueshifts of the
cores of their \oiii\ emission lines than
BLS1 galaxies, including a number of ``blue outliers'' with
extreme outflows (K08).
Several of these blue outliers
are located in the high Eddington ratio and high luminosity
corner of the EV1-EV2 diagram (Figure\,6).
Interestingly, broad absorption line (BAL) QSOs,
which show high outflow velocities in their absorption-line
spectra (e.g., Weymann et al. 1991), fall into a similarly extreme
locus (e.g., Boroson 2002) as the blue outliers of our sample,
therefore pointing to possible links between these two
source classes. Possibly, they bridge a gap between
NLS1 galaxies and BAL QSOs (see the discussion 
of, e.g., Brandt \& Gallagher 2000 on general
similarities between NLS1 galaxies and BAL quasars).
UV spectroscopy is needed to see whether the BAL fraction
is high in these extreme blue outliers.

\subsection{Links between central engine and host properties}

Based on the correlation analyses we carried out, 
the strongest correlations among the directly
measured parameters of our sample are
between FWHM(\hbb), R4570 and R5007, confirming
previous studies (e.g., Boroson \& Green 1992, Grupe 2004).
This confirmation is of interest, because
few independent large NLS1-BLS1 samples exist, for which a rigorous
correlation analysis has been performed. Samples had different
selection effects (e.g., soft X-ray selection), and therefore
new NLS1 samples which were selected differently from
previous ones are important to check
the persistency of trends.

More importantly, we have shown that
density is {\em as important an ingredient in correlation space 
as Eddington ratio}. 
Including density is of particular interest,
because the NLR density is representative of the interstellar
medium (ISM) of the host galaxy,
and this study therefore links the NLS1 central engine properties
with the host galaxy. Host properties are otherwise
not easily accessible, and have only been studied
for a few samples of NLS1 galaxies, focusing on the
host morphology
(Crenshaw et al. 2003, Deo et al. 2006, Ohta et al. 2007).
Especially, host properties of NLS1 galaxies are an independent
indicator of their evolutionary stage
(Mathur et al. 2011, Orban de Xivry et al. 2011).

Several factors can affect the density of the ISM of the host galaxy.
Winds and outflows on the one hand (e.g., Schiano 1986, Kaiser et al. 2000),
(bar-driven) inflows on the other hand (e.g., Shlosman et al. 1990,
Riffel et al. 2008).
We comment on each of them in turn.

Winds/outflows are seen in nearby Seyfert galaxies (e.g., Kraemer et al. 2008,
Wang et al. 2010) and in NLS1 galaxies in particular (e.g.,
Komossa et al. 2008, Zhang et al. 2011).
These can be disk-wind driven, or also be powered by radio jets
(Morganti et al. 2010).
The latter effect is unlikely to dominate in NLS1 galaxies.
Even though a small
fraction of them shows evidence for relativistic
jets in radio and gamma-rays
(e.g., Zhou et al. 2003, Yuan et al. 2008, Komossa et al. 2006,
Abdo et al. 2009, Foschini 2011),
the majority of them is less radio-loud, on average, than
BLS1 galaxies (Komossa et al. 2006), and rarely
shows widely extended radio emission.
Eddington-driven winds are more likely (e.g., Grupe 2004), given the
high values of $L/L_{\rm Edd}$ of NLS1 galaxies. These winds
may extend into the inner NLR (Proga et al. 2008), and possibly further.
Galaxy merger simulations predict large-scale outflows
(e.g., di Matteo et al. 2005),
but the merger fraction
among NLS1 galaxies is still poorly constrained (e.g., Krongold et al. 2001),
and we do not find a strong excess of pairing among our sample.

On the other hand, recent studies indicate an excess of bars
in NLS1 galaxies (Crenshaw et al. 2003, Ohta et al. 2007),
and bar-driven instabilities
might work at replenishing the NLR with (low-density) gas.
Indeed, theoretical studies have shown that bars are efficient to
transport large amounts of gas inward
(e.g., Shlosman et al. 1990).  
Simulations have further shown that strong bars drive the formation of, and
maintain grand-design nuclear dust spiral structures in
the central kiloparsec of galaxies (e.g., Maciejewski 2004a,b).
The gas loses its angular momentum to the bar via nuclear
spiral shocks and is eventually concentrated in a narrow nuclear ring
if it follows closely the periodic orbits in the bar
(e.g., Patsis \& Athanassoula 2000, Maciejewski et al. 2002).
Regarding NLS1 galaxies, uninterrupted asymmetries able to drive
gas inwards all the way from a few kpc down to a few tens of pc
(e.g., Orban de Xivry et al. 2011) are detected.
In particular, NLS1 galaxies show a
higher fraction of grand-design dust spirals within $\sim 1\,{\rm kpc}$
and stellar nuclear rings than BLS1 galaxies (Deo et al. 2006).
Bar instabilities can also drive the formation of pseudo-bulges by
internal secular processes (e.g. Kormendy \& Kennicutt 2004).
Such secular processes in NLS1 galaxy have been inferred based
on bulge-disc decompositions (Ryan et al. 2007, Mathur et al. 2011,
Orban de Xivry et al. 2011).

Well-resolved host images are not yet available for the bulk of
the NLS1 galaxies of our sample, for further distinguishing
between the two scenarios discussed above, but could be obtained in the
future with HST.
Combining other host galaxy properties
(structure, presence and properties of bars, stellar populations)
with emission-line properties into correlation analyses
will be another important future step.

\section{Summary and conclusions}

We have studied the correlations among the measured and derived
properties of a sample of NLS1 and BLS1 galaxies
on the basis of a principle component analysis.
For the first time, we have involved the density of
the narrow-line region (measured by the density-sensitive
ratio of the two emission lines \sii\ $\lambda\lambda$6716,6713)
in such an analysis.
A nearest-neighbor analysis of the large-scale environment of NLS1
galaxies was also performed. 
Our main results can be summarized as follows: 

\begin{itemize}
\item
We have found that, among the parameters measured in this sample, 
the density of the narrow-line region 
is a key element of Eigenvector 1 (EV1),
as important as the Eddington ratio \LLedd.
This is of particular interest, because it links the
central engine and the host properties.

\item
Apart from this new finding, we also confirm several previously known trends,
especially the strong correlations involving the line width of \hb,
and the strength of the \feii\ complex and \oiii\ $\lambda$5007 emission.
In addition to density, \LLedd\ plays a significant role
in affecting EV1 of our sample, while 
EV2 is related to luminosity.
NLS1 and BLS1 galaxies are well separated in EV1 space,
while they are not distinguished in EV2 space.

\item 
Accretion-driven winds on the one hand, or bar-driven inflows
on the other hand, may play a role in explaining the links between
the immediate vicinity of the supermassive black hole (SMBH)
on the one hand (as traced by
Eddington ratio and broad-line width, for instance),
and the properties (especially density) of the host galaxy
on the other hand.

\item
Several galaxies with strong blueshifts of \oiii\  (``blue outliers'')
lie in the high Eddington ratio
and high luminosity regime in the EV1-EV2 space,
possibly sharing these properties with broad absorption line QSOs,
and therefore suggesting possible connections between these two source classes.

\item
To the first order, the NLS1 and BLS1 galaxies of our sample are
found to reside in similar large-scale environments.
While this needs to be confirmed with larger samples, it
tentatively indicates the lack of an excess of mergers among
NLS1 galaxies.

\end{itemize}

Large-sample analyses of NLS1 galaxies and their remarkable properties
are an important new approach in our understanding of black hole growth,
accretion modes, feeding and feedback, and of aspects
of galaxy -- SMBH co-evolution and the role of secular evolution.

\acknowledgments
We thank T. Wang for discussions and comments. 
This work is supported by the National Natural Science Foundation of China 
(Grant No. 10873017) and National Basic Research Program of China - 973
program (Grant No. 2009CB824800). 
D.X. thanks the Max-Planck-Institut f\"ur extraterrestrische Physik and
the Max-Planck-Gesellschaft for support and hospitality. 
S.K. thanks the Aspen Center for Physics for their hospitality.
H.Z. acknowledges support from the Alexander von Humboldt Foundation,
from NSFC (grant NSF-10533050), and from program 973 (No. 2007CB815405).
C.L. acknowledges support from NSFC (No. 11173045),
Shanghai Pujiang Program (No. 11PJ1411600) and the
CAS/SAFEA International Partnership Program for Creative
Research Teams (KJCX2-YW-T23).
W.Y. acknowledges support from NSFC (No. 11033007). 
This research has made use of the SDSS data base, and of the 
NASA/IPAC Extragalactic Database (NED) which is operated 
by the Jet Propulsion Laboratory, California Institute of
Technology, under contract with the National Aeronautics and 
Space Administration. Funding for the SDSS and SDSS-II has 
been provided by the Alfred P. Sloan Foundation,
the Participating Institutions, the National Science Foundation,
the U.S. Department of Energy, the National Aeronautics and
Space Administration, the Japanese Monbukagakusho, 
the Max Planck Society, and the Higher Education Funding Council for England.
The SDSS Web Site is http://www.sdss.org/.
The SDSS is managed by the Astrophysical Research Consortium
for the Participating Institutions. The Participating Institutions are 
the American Museum of Natural History, Astrophysical Institute Potsdam,
University of Basel, University of Cambridge, Case Western Reserve University, 
University of Chicago, Drexel University, Fermilab, the Institute for 
Advanced Study, the Japan Participation Group, Johns Hopkins University, 
the Joint Institute for Nuclear Astrophysics, the Kavli Institute for
Particle Astrophysics and Cosmology, the Korean Scientist Group, 
the Chinese Academy of Sciences (LAMOST), Los Alamos National Laboratory, 
the Max-Planck-Institute for Astronomy (MPIA), the Max-Planck-Institute 
for Astrophysics (MPA), New Mexico State University, Ohio State University, 
University of Pittsburgh, University of Portsmouth, Princeton University, 
the United States Naval Observatory, and the University of Washington.

\clearpage
\setlength{\voffset}{10mm}
\begin{deluxetable}{llcccccccccccccc}
\tabcolsep 1.2mm
\rotate
\tabletypesize{\tiny}
\tablecaption{Key properties of the NLS1 and BLS1 galaxies}
\tablewidth{0pt}
\tablehead{
\colhead{coordinates\,(J2000)} & \colhead{common name} & \colhead{$z$} & \colhead{$M_{\rm i}$} & \colhead{FWHM(\hbb)} & \colhead{R5007} & \colhead{R4570}
& \colhead{FWHM([S\,II])} & \colhead{FWHM([O\,III]$_c$)} & \colhead{$v_{\rm [OIII]_c}$} & \colhead{R$_{\rm [SII]}$} & \colhead{$n_{\rm e}$} & \colhead{$f_{\rm 5100}$}& \colhead{log\,$\lambda L_{\rm 5100}$} & \colhead{log\,$L/L_{\rm Edd}$} & \colhead{log\,$M_{\rm BH}/{\rm M}_{\odot}$}  \\
\colhead{(1)\tablenotemark{a}} & \colhead{(2)} & \colhead{(3)} & \colhead{(4)} & \colhead{(5)} & \colhead{(6)} & \colhead{(7)}  & \colhead{(8)}& \colhead{(9)}  & \colhead{(10)} & \colhead{(11)} & \colhead{(12)} & \colhead{(13)} & \colhead{(14)} & \colhead{(15)} & \colhead{(16)}}
\startdata
001903.17$+$000659.2 & SDSSJ00190+0006 & 0.073 & -20.1 & 2870 & 0.7 & 0.2 & 200 &160 & -10 & 0.99 & 630 & 15.9 & 43.0 & -0.9 & 6.7 \\
002305.04$-$010743.4 & 0020-0124       & 0.166 & -21.2 & 1330 & 0.3 & 1.2 & 180 &320 & 90  & 1.31 & 110 & 13.7 & 43.6 & -0.03 & 6.5 \\
002752.39$+$002615.8 & Q0025+0009      & 0.205 & -22.2 & 2180 & 0.5 & 0.6 & 340 &310 & -20 & 1.24 & 190 & 21.4 & 44.0 & -0.3 & 7.1 \\
003238.20$-$010035.2 & SDSSJ00326-0100 & 0.092 & -20.8 & 1760 & 0.6 & 0.6 & 180 &110 & 40  & 1.18 & 270 & 23.0 & 43.3 & -0.4 & 6.5 \\
003711.00$+$002128.0 & SDSSJ00371+0021 & 0.235 & -22.1 & 1920 & 0.6 & 0.6 & 150 &300 & 90  & 1.38 & 40 & 17.8 & 44.0  & -0.2 & 7.0 \\
003847.97$+$003457.4 & SDSSJ00387+0034 & 0.081 & -21.0 & 6170 & 0.5 & 0.1 & 370 &310 & -60 & 1.00 & 610 & 29.1 & 43.3 & -1.4 & 7.6 \\
010939.02$+$005950.3 & RXJ01097+0059   & 0.093 & -21.0 & 2730 & 1.3 & 0.1 & 340 &260 & 30  & 1.03 & 530 & 20.3 & 43.3 & -0.8 & 6.9 \\
011448.68$-$002946.1 & SDSSJ01148-0029 & 0.034 & -19.1 & 2880 & 0.6 & 0.4 & 200 &200 & -10 & 1.47 & -- & 29.8  & 42.6 & -1.0 & 6.5 \\
011703.58$+$000027.4 & 2E0114-0015     & 0.046 & -20.0 & 2390 & 0.2 & 0.4 & 280 &310 & -10 & 1.12 & 360 & 39.2 & 43.0 & -0.7 & 6.5 \\
011929.06$-$000839.7 & NGC450-86       & 0.090 & -20.8 & 1220 & 0.5 & 0.9 & 300 &380 & 220 & 1.23 & 200 & 23.8 & 43.3 & -0.04 & 6.2 \\
012159.82$-$010224.4 & IIZw1           & 0.054 & -20.9 & 3280 & 1.1 & 0.3 & 310 &280 & 90  & 1.06 & 470 & 79.2 & 43.4 & -0.9 & 7.1\\
013521.68$-$004402.2 & RXJ01354-0043   & 0.099 & -21.0 & 1710 & 1.1 & 0.5 & 250 &620 & 240 & 1.33 & 90  & 21.5 & 43.4 & -0.3 & 6.5  \\
013527.85$-$004448.0 & SDSSJ01354-0044 & 0.080 & -20.7 & 2890 & 0.2 & 0.1 & 330 &340 & 0   & 1.12 & 360 & 17.7 & 43.1 & -0.9 & 6.8  \\
013940.99$-$010944.4 & SDSSJ01396-0109 & 0.194 & -21.7 & 1690 & 1.0 & 0.3 & 250 &310 & 80  & 1.24 & 190 & 15.3 & 43.8 & -0.2 & 6.8 \\
014644.82$-$004043.2 & NPM1G-00.0070   & 0.083 & -20.5 & 1350 & 0.4 & 0.4 & 220 &140 & 90  & 1.24 & 190 & 18.2 & 43.1 & -0.2 & 6.2 \\
014951.66$+$002536.5 & SDSSJ01498+0025 & 0.252 & -21.7 & 2080 & 0.7 & 0.6 & 180 &170 & 20  & 1.20 & 240 & 11.6 & 43.9 & -0.3 & 7.0 \\
015652.43$-$001222.0 & SDSSJ01568-0012 & 0.163 & -21.4 & 1510 & 0.5 & 1.1 & 330 &280 & 0   & 1.27 & 150 & 17.9 & 43.7 & -0.1 & 6.6  \\
020615.99$-$001729.2 & MARK1018        & 0.043 & -21.6 & 4040 & 0.5 & 0.02 & 340& 390 & 0  & 1.03 & 530 & 251  & 43.7 & -0.95 & 7.5 \\
021359.78$+$004226.8 & RXJ02139+0042   & 0.182 & -22.8 & 5670 & 1.8 & 0.00 & 470 & 370 & 0  & 1.18 & 270 & 43.6 & 44.2 & -1.1 & 8.1 \\
022756.28$+$005733.1 & SDSSJ02279+0057 & 0.128 & -20.2 & 1110 & 0.6 & 0.4 & 130 &130 & 60  & 1.23 & 200 & 6.4  & 43.0 & -0.05 & 5.9 \\
022841.48$+$005208.6 & SDSSJ02286+0052 & 0.186 & -21.6 & 1170 & 0.7 & 0.9 & 270 &390 & 90  & 1.36 & 60 & 14.2  & 43.7 & 0.1 & 6.4 \\
025646.96$+$011349.4 & S0254+0101      & 0.177 & -22.6 & 2590 & 0.6 & 0.3 & 570 &470 & 30  & 1.07 & 450 & 34.2 & 44.0 & -0.5 & 7.3 \\
030124.26$+$011023.0 & SDSSJ03014+0110 & 0.072 & -20.7 & 2940 & 0.4 & 1.0 & 250 &360 & 110 & 1.19 & 250 & 33.5 & 43.3 & -0.8 & 6.9 \\
030144.20$+$011530.9 & RXJ03017+0115   & 0.075 & -21.1 & 3450 & 0.6 & 0.4 & 340 &300 & 40  & 1.13 & 340 & 43.2 & 43.4 & -0.9 & 7.2 \\
030417.78$+$002827.4 & KUG0301+002     & 0.045 & -19.7 & 1200 & 0.6 & 0.1 & 220 &110 & 20  & 1.20 & 240 & 31.6 & 42.9 & -0.2 & 5.9 \\
030639.57$+$000343.2 & SDSSJ03066+0003 & 0.107 & -21.8 & 1840 & 0.6 & 0.2 & 330 &380 & -30 & 1.20 & 240 & 35.8 & 43.6 & -0.3 & 6.8 \\
031027.83$-$004950.8 & KUV03079-0101   & 0.080 & -22.4 & 2820 & 0.3 & 0.5 & 270 &130 & -10 & 1.19 & 250 & 150  & 44.0 & -0.5 & 7.4 \\
032213.90$+$005513.5 & KUV03197+0045   & 0.185 & -23.8 & 2490 & 0.3 & 0.3 & 300 &270 & 50  & 1.05 & 490 & 110.8 & 44.6 & -0.2 & 7.7 \\
032337.65$+$003555.7 & SDSSJ03236+0035 & 0.215 & -22.8 & 1740 & 0.2 & 0.5 & 190 &270 & 110 & 1.08 & 430 & 38.2 & 44.3 & -0.04 & 7.1 \\
032606.75$+$011429.9 & SDSSJ03261+0114 & 0.127 & -21.3 & 1230 & 0.5 & 0.9 & 330 &530 & 180 & 1.13 & 340 & 16.4 & 43.5 & -0.01 & 6.3 \\
032729.88$-$005958.5 & SDSSJ03274-0059 & 0.134 & -21.5 & 5690 & 0.8 & 0.3 & 290 &290 & 20  & 1.04 & 510 & 18.2 & 43.5 & -1.3 & 7.7 \\
034131.95$-$000933.0 & SDSSJ03415-0009 & 0.223 & -21.4 & 1480 & 0.6 & 0.7 & 200 &170 & 10  & 1.36 &  60 &  8.9 & 43.7 & -0.1 & 6.6 \\
075245.60$+$261735.8 & RXJ07527+2617   & 0.082 & -21.4 & 1600 & 0.3 & 0.8 & 290 &210 & 30  & 1.06 & 470 & 55.1 & 43.6 & -0.2 & 6.6 \\
083949.65$+$484701.5 & NPM1G+48.0114   & 0.039 & -19.8 & 1290 & 1.2 & 0.2 & 220 &190 & 20  & 1.19 & 250 & 50.4 & 42.9 & -0.2 & 6.0 \\
091313.73$+$365817.3 & RXJ09132+3658   & 0.107 & -21.0 & 1680 & 1.0 & 0.5 & 160 &350 & 150 & 1.38 & 40 & 21.3 & 43.4  & -0.3 & 6.5 \\
092247.03$+$512038.0 & SBS0919+515     & 0.160 & -22.3 & 1250 & 0.3 & 1.3 & 160 &720 & 430 & 1.42 & 10 & 38.5 & 44.0 & 0.2 & 6.7 \\
100405.00$-$003253.4 & SDSSJ10040-0032 & 0.289 & -22.6 & 1410 & 0.4 & 2.0 & 150 &140 & 40  & 1.21 & 230 & 28.8 & 44.4 & 0.2 & 7.0 \\
101645.11$+$421025.5 & RXJ10167+4210   & 0.055 & -21.2 & 1690 & 0.2 & 0.8 & 270 &250 & 20  & 1.23 & 200 & 92.4 & 43.5 & -0.3 & 6.6 \\
102434.72$+$555626.5 & SBS1021+561     & 0.197 & -22.4 & 1660 & 0.6 & 0.7 & 200 &240 & 50  & 1.20 & 240 & 29.9 & 44.1 & -0.1 & 7.0 \\
102448.57$+$003538.0 & SDSSJ10248+0035 & 0.095 & -20.8 & 1990 & 0.6 & 0.3 & 350 &230 & 0   & 1.26 & 160 & 17.1 & 43.2 & -0.5 & 6.6 \\
102531.29$+$514034.9 & MARK142         & 0.045 & -20.6 & 1530 & 0.2 & 0.9 & 180 &250 & 50  & 1.28 & 140 & 97.2 & 43.3 & -0.2 & 6.4 \\
104210.03$-$001814.7 & SDSSJ10421-0018 & 0.115 & -19.9 & 1150 & 0.3 & 0.9 & 170 &210 & 50  & 1.33 & 90 & 6.8   & 43.0 & -0.1 & 5.9 \\
104230.14$+$010223.7 & SDSSJ10425+0102 & 0.115 & -20.8 & 1430 & 1.1 & 0.5 & 260 &300 & 90  & 1.18 & 270 & 13.4 & 43.3 & -0.2 & 6.3 \\
104332.88$+$010108.8 & SDSSJ10435+0101 & 0.072 & -20.4 & 2530 & 1.0 & 0.1 & 230 &200 & 0   & 1.19 & 250 & 18.4 & 43.0 & -0.8 & 6.6 \\
112941.94$+$512050.7 & SBS1126+516     & 0.234 & -23.2 & 1880 & 0.6 & 0.3 & 190 &250 & 60  & 1.17 & 280 & 52.1 & 44.5 & -0.04 & 7.3 \\
115023.59$+$000839.1 & Q1147+0025      & 0.127 & -21.1 & 1310 & 0.5 & 0.6 & 210 &150 & 0   & 0.97 & 690 & 16.7 & 43.5 & -0.1 & 6.4 \\
115235.00$-$000542.8 & Q1150+0010      & 0.129 & -21.6 & 3140 & 0.4 & 0.4 & 280 &190 & 10  & 1.14 & 330 & 27.1 & 43.7 & -0.7 & 7.3 \\
115533.50$+$010730.6 & SDSSJ11555+0107 & 0.197 & -21.9 & 1510 & 0.3 & 0.7 & 150 &780 & 330 & 1.23 & 200 & 17.1 & 43.8 & -0.05 & 6.7 \\
115758.73$-$002220.8 & QUESTJ1157-0022 & 0.260 & -23.8 & 4670 & 0.6 & 0.3 & 360 &280 & -10 & 1.20 & 240 & 93.4 & 44.8 & -0.7 & 8.4 \\
120226.76$-$012915.3 & IRAS11598-0112  & 0.150 & -22.4 & 1460 & 0.5 & 2.7 & 250 &340 & 170 & 1.31 & 110 & 47.6 & 44.1 &  0.1 & 6.8 \\
121549.45$+$544224.0 & SBS1213+549A    & 0.150 & -22.9 & 1240 & 0.6 & 1.2 & 310 &320 & 110 & 1.27 & 150 & 73.4 & 44.2 &  0.3 & 6.8 \\
124324.22$+$010028.1 & SDSSJ12434+0100 & 0.090 & -20.6 & 5180 & 0.4 & 0.2 & 230 &240 & -10 & 0.94 & 770 & 16.5 & 43.2 & -1.3 & 7.3 \\
124623.00$+$002839.9 & SDSSJ12463+0028 & 0.088 & -20.6 & 2230 & 0.9 & 0.3 & 220 &210 & 40  & 1.14 & 330 & 17.6 & 43.2 & -0.6 & 6.6 \\
124635.25$+$022208.8 & PG1244+026      & 0.048 & -20.5 & 1200 & 0.6 & 0.8 & 230 &300 & 150 & 1.10 & 400 & 82.0 & 43.3 & -0.03 & 6.2 \\
130023.22$-$005429.8 & UM534           & 0.122 & -21.3 & 1240 & 0.7 & 0.6 & 250 &150 & 70  & 1.09 & 410 & 20.9 & 43.5 &  0.01  & 6.3 \\
130713.25$-$003601.6 & SDSSJ13072-0036 & 0.170 & -22.5 & 2240 & 0.9 & 0.3 & 370 &310 & -30 & 1.01 & 580 & 40.6 & 44.1 & -0.3 & 7.2 \\
131750.32$+$601041.0 & SBS1315+604     & 0.137 & -22.1 & 2660 & 1.5 & 0.1 & 380 &340 & 20  & 1.20 & 240 & 31.3 & 43.8 & -0.6 & 7.2 \\
132135.33$-$001305.8 & SDSSJ13215-0013 & 0.082 & -21.2 & 3300 & 1.0 & 0.3 & 240 &300 & 130 & 1.11 & 380 & 38.5 & 43.5 & -0.9 & 7.1 \\
134351.07$+$000434.8 & SDSSJ13438+0004 & 0.074 & -20.7 & 1960 & 0.7 & 0.4 & 200 &190 & 10  & 0.99 & 630 & 29.2 & 43.2 & -0.4 & 6.5 \\
134459.45$-$001559.5 & Q1342-000       & 0.245 & -23.0 & 2450 & 1.0 & 0.2 & 410 &240 & -30 & 1.26 & 160 & 41.7 & 44.4 & -0.3 & 7.5 \\
135516.55$+$561244.7 & SBS1353+564     & 0.122 & -22.4 & 1340 & 1.6 & 0.7 & 490 &440 & 50  & 0.94 & 770 & 60.2 & 44.0 &  0.1  & 6.7 \\
143030.22$-$001115.1 & SDSSJ14305-0011 & 0.103 & -20.1 & 1510 & 1.3 & 0.5 & 170 &160 & 30  & 1.23 & 200 & 8.4  & 43.0 & -0.3 & 6.2 \\
143039.31$+$493539.0 & CSO661          & 0.204 & -22.7 & 2480 & 0.3 & 0.5 & 260 &230 & 30  & 1.19 & 250 & 38.9 & 44.2 & -0.4 & 7.4 \\
143847.54$-$000805.4 & SDSSJ14387-0008 & 0.104 & -20.9 & 3570 & 0.2 & 0.4 & 200 &210 & -20 & 1.01 & 580 & 15.1 & 43.2 & -1.0 & 7.1 \\
144130.11$+$592801.7 & RXJ14414+5928   & 0.134 & -21.2 & 1600 & 0.3 & 0.8 & 180 &250 & 30  & 1.36 & 60 & 16.8  & 43.5 & -0.2 & 6.6 \\
144735.26$-$003230.5 & SDSSJ14475-0032 & 0.217 & -21.6 & 1470 & 0.3 & 0.6 & 140 &140 & 30  & 1.20 & 240 & 12.1 & 43.8 & -0.1 & 6.7 \\
144932.71$+$002236.3 & SDSSJ14495+0022 & 0.081 & -20.5 & 1240 & 0.7 & 0.8 & 230 &140 & 40  & 1.30 & 120 & 18.0 & 43.1 & -0.1 & 6.1 \\
145123.02$-$000625.9 & SDSSJ14513-0006 & 0.139 & -21.1 & 2610 & 0.8 & 0.1 & 200 &280 & -10 & 1.23 & 200 & 12.3 & 43.4 & -0.7 & 6.9 \\
145143.30$+$524127.4 & RXJ14517+5241   & 0.206 & -22.0 & 1970 & 0.5 & 0.2 & 220 &290 & 60  & 1.04 & 510 & 18.3 & 43.9 & -0.3 & 7.0 \\
145631.65$-$001114.2 & SDSSJ14565-0011 & 0.132 & -21.8 & 6220 & 0.5 & 0.1 & 220 &230 & -20 & 0.87 & 1030& 25.3 & 43.7 & -1.3 & 7.8 \\
151956.57$+$001614.6 & SDSSJ15199+0016 & 0.114 & -21.5 & 1730 & 0.4 & 0.4 & 320 &340 & 90  & 1.07 & 450 & 26.3 & 43.6 & -0.3 & 6.7 \\
152628.20$-$003809.4 & SDSSJ15264-0038 & 0.123 & -21.5 & 2240 & 0.3 & 0.8 & 120 &170 & 20  & 1.25 & 180 & 27.4 & 43.6 & -0.5 & 6.9 \\
153732.62$+$494247.8 & SBS1536+498     & 0.280 & -22.2 & 1290 & 0.5 & 1.0 & 120 &140 & 30  & 1.29 & 130 & 21.6 & 44.2 & 0.2 & 6.9 \\
153911.17$+$002600.8 & SDSSJ15391+0026 & 0.265 & -22.6 & 1450 & 0.7 & 0.4 & 220 &380 & 110 & 1.25 & 180 & 28.2 & 44.3 & 0.1 & 7.0 \\
155922.19$+$270338.9 & 2E1557+2712     & 0.064 & -20.7 & 2500 & 0.5 & 0.3 & 250 &200 & -10 & 1.23 & 200 & 35.7 & 43.2 & -0.7 & 6.7 \\
161809.38$+$361957.8 & RXJ16181+3619   & 0.034 & -19.1 & 1150 & 0.6 & 0.9 & 120 &130 & 20  & 1.43 & $<$10& 35.0& 42.7 & -0.2 & 5.7 \\
161951.31$+$405847.2 & KUG1618+410     & 0.038 & -19.2 & 1750 & 1.1 & 0.2 & 280 &210 & 20  & 1.16 & 300 & 28.5 & 42.7 & -0.6 & 6.1 \\
163323.58$+$471859.0 & RXJ16333+4718   & 0.116 & -21.7 & 1260 & 1.2 & 0.9 & 340 &250 & 40  & 1.17 & 280 & 38.0 & 43.7 & 0.1 & 6.5 \\
164907.64$+$642422.3 & SDSSJ16491+6424 & 0.184 & -21.1 & 1360 & 0.5 & 0.8 & 160 &320 & 10  & 1.41 & 20 & 9.4 & 43.5   & -0.1 & 6.4 \\
165408.16$+$392533.4 & EXO1652.4+3930  & 0.069 & -20.4 & 1280 & 0.2 & 0.6 & 250 &120 & 50  & 1.11 & 380 & 25.1 & 43.1 & -0.2 & 6.1 \\
165658.38$+$630051.1 & SDSSJ16569+6300 & 0.169 & -21.1 & 1720 & 0.2 & 0.7 & 160 &210 & 20  & 1.15 & 310 & 10.8 & 43.5 & -0.3 & 6.6 \\
170328.97$+$614110.0 & SDSSJ17034+6141 & 0.077 & -21.5 & 4000 & 0.6 & 0.2 & 260 &430 & 60  & 1.16 & 300 & 55.1 & 43.6 & -1.0 & 7.4 \\
170546.91$+$631059.1 & SDSSJ17057+6310 & 0.119 & -20.9 & 1930 & 0.2 & 0.7 & 250 &230 & 80  & 1.37 & 50 & 14.9 & 43.4  & -0.4 & 6.6 \\
170812.29$+$601512.6 & SDSSJ17082+6015 & 0.145 & -21.0 & 1070 & 1.0 & 0.5 & 280 &260 & 60  & 1.25 & 180 & 10.8 & 43.4 & 0.1 & 6.1 \\
171411.63$+$575834.1 & SBS1713+580     & 0.093 & -22.2 & 2130 & 0.4 & 0.7 & 360 &300 & 6   & 1.13 & 340 & 88.7 & 43.9 & -0.3 & 7.1 \\
171550.49$+$593548.8 & SDSSJ17158+5935 & 0.066 & -20.0 & 3240 & 0.7 & 0.2 & 340 &270 & 40  & 1.14 & 330 & 15.8 & 42.9 & -1.0 & 6.7 \\
171829.01$+$573422.4 & SDSSJ17184+5734 & 0.101 & -21.3 & 1760 & 0.4 & 0.7 & 110 &470 & 150 & 1.40 & 30 & 26.9 & 43.5 & -0.3 & 6.6 \\
172032.29$+$551330.3 & SDSSJ17205+5513 & 0.273 & -22.7 & 3490 & 0.6 & 0.2 & 250 &250 & -10 & 1.20 & 240 & 33.7 & 44.4 & -0.6 & 7.8 \\
172533.07$+$571645.6 & SDSSJ17255+5716 & 0.066 & -20.5 & 4870 & 1.4 & 0.04& 280 &210 & 20 & 1.26 & 160 & 23.1 & 43.0 & -1.3 & 7.2 \\
173107.87$+$620026.1 & SDSSJ17311+6200 & 0.069 & -20.8 & 3860 & 1.1 & 0.4 & 350 &310 & 20  & 1.07 & 450 & 30.5 & 43.2 & -1.1 & 7.1 \\
221918.53$+$120753.2 & IIZw177         & 0.081 & -21.3 & 1180 & 0.5 & 1.2 & 170 &200 & 40  & 1.29 & 130 & 46.6 & 43.5 & 0.1 & 6.3 \\
232328.00$+$002032.9 & SDSSJ23234+0020 & 0.120 & -21.2 & 3890 & 0.5 & 0.3 & 410 &300 & 10  & 1.11 & 380 & 18.4 & 43.5 & -1.0 & 7.3 \\
233032.95$+$000026.4 & SDSSJ23305+0000 & 0.123 & -21.3 & 1220 & 0.7 & 0.8 & 410 &280 & 80  & 1.15 & 310 & 18.5 & 43.5 & 0.01 & 6.3 \\
234725.30$-$010643.7 & SDSSJ23474-0106 & 0.182 & -21.8 & 1830 & 0.3 & 0.4 & 320 &180 & 40  & 1.21 & 230 & 18.2 & 43.8 & -0.2 & 6.9
\enddata
\tablenotetext{a}
{
columns from left to right: 
(1) coordinates in RA (h,m,s) and DEC (d,m,s), 
(2) galaxy name, (3) redshift, (4) SDSS absolute i magnitude, 
(5) FWHM of the broad component of H$\beta$ in \kms,
(6) ratio of total [O III] over total H$\beta$ emission, 
(7) ratio of Fe\,II\,4570 over total H$\beta$ emission,
(8) FWHM of [S II] in \kms, (9) FWHM of the core of [O III] in \kms, 
(10) velocity shift (blueshift) of the core of [O III] with respect 
to [S II] in \kms.
(11) intensity ratio of [S II] $\lambda6716/\lambda6716$, 
(12) NLR electron density in cm$^{-3}$, 
(13) flux density at 5100\,\AA\, restframe in 
$10^{-17}$~erg~s$^{-1}$~cm$^{-2}$~\AA$^{-1}$, 
(14) $\log$ of the monochromatic luminosity at 5100\,\AA, 
(15) $\log$ of the Eddington ratio, (16) $\log$ of the black hole mass.
}
\label{table1}
\end{deluxetable}

\setlength{\hoffset}{0mm}
\begin{deluxetable}{llcc}
\tabcolsep 1.2mm
\tablecaption{Sample properties}
\tablewidth{0pt}
\tablehead{
\colhead{Prop.} &\colhead{class} &\colhead{range} &\colhead{average}
}
\startdata
$\log M_{\rm BH}/M_{\odot}$  & NLS1 & [5.7,7.3] & 6.5  \\
                             & BLS1 & [6.5,8.4] & 7.2  \\
$\log L/L_{\rm Edd}$         & NLS1 &[-0.6,0.3] &-0.1  \\
                             & BLS1 &[-1.4,-0.3]&-0.8  \\
FWHM(${\rm {[SII]}}$)          & NLS1 &[110,490]  & 230  \\
                             & BLS1 &[120,570]  & 300  \\
FWHM(${\rm {[OIII]_c}}$)       & NLS1 &[110,780]  & 280   \\
                             & NLS1\tablenotemark{a} &[110,440] & 210 \\
                             & BLS1 &[140,470]  & 270  \\
                             & BLS1\tablenotemark{a} &[140,470] &270 \\ 
R4570                        & NLS1 &[0.1,2.7]  & 0.7 \\
                             & BLS1 &[0.0,1.0]  & 0.3  \\
R5007                        & NLS1 &[0.2,1.6]  & 0.6  \\
                             & BLS1 &[0.2,1.8]  & 0.7  \\
$v_{\rm_{[OIII]_c}}$         & NLS1 &[-30,430]  & 80   \\
                             & BLS1 &[-60,130]  & 10   \\
R${\rm_{[SII]}}$             & NLS1 &[0.94,1.43]&1.22  \\
                             & BLS1 &[0.87,1.47]&1.13  \\
$n_{\rm e}$                  & NLS1 &[10,770]\tablenotemark{b} &200  \\
                             & BLS1 &[150,1030] &340  \\
\enddata
\tablenotetext{a}
{Excluding galaxies with high velocity shifts (i.e., $v_{\rm_{[OIII]_c}} > 75$ \kms).}  
\tablenotetext{b}
{Two objects, RXJ16181+3619 and SDSSJ01148-0029, have 
\n $<$ 10\,\cm, and were not included in the analysis involving \n.}
\end{deluxetable}

\clearpage
\setlength{\hoffset}{-10mm}
\begin{deluxetable}{lccccccccccc}
\tabcolsep 1.2mm
\tabletypesize{\scriptsize}
\tablewidth{0pt}
\tablecaption{Spearman rank order correlation\tablenotemark{a}}
\tablehead{
\colhead{Prop.} & \colhead{FWHM$_{\rm H\beta_b}$} & \colhead{R5007} & 
\colhead{R4570} & \colhead{FWHM$_{\rm [SII]}$} & \colhead{FWHM$_{\rm [OIII]_c}$} & 
\colhead{$v_{\rm_{[OIII]_c}}$} & \colhead{R$_{\rm [SII]}$} & \colhead{$n_{\rm e}$} & 
\colhead{$\lambda L_{\rm 5100}$} & \colhead{$L/L_{\rm Edd}$} & 
\colhead{$M_{\rm BH}$} \\
\colhead{} & \colhead{(1)} & \colhead{(2)} & \colhead{(3)} & \colhead{(4)} 
& \colhead{(5)} & \colhead{(6)} & \colhead{(7)} & \colhead{(8)} 
& \colhead{(9)} & \colhead{(10)} & \colhead{(11)}
}
\startdata
(1) FWHM$_{\rm H\beta_b}$ &  all & 0.09    &{\bf -0.70}   &{\bf 0.38} &  0.15       &{\bf -0.52} &{\bf -0.43} &{\bf 0.46}   &   0.04 &{\bf -0.91}   &{\bf 0.79}\\
                       & NLS1s&-0.10    &{\bf -0.45}   &   0.02    &  0.12       &   -0.15    &   -0.11    &     0.11    &   0.22 &{\bf -0.67}   &{\bf 0.59}\\
                       & BLS1s& 0.14    &{\bf -0.50}   &   0.15    &  0.24       &   -0.07    &   -0.37    &     0.38    &  -0.13 &{\bf -0.85}   &{\bf 0.51}\\
(2) R5007               & 0.09 & all     &{\bf -0.38}   &{\bf 0.29} &  0.10       &-0.04       &   -0.07    &  0.07       &  -0.10 & -0.14        &   0.01   \\
                       &-0.10 & NLS1s   & {\bf -0.35}  &   0.30    &  0.04       &   -0.08    &   -0.09    &     0.09    &  -0.16 &  0.01        &  -0.20   \\
                       & 0.14 & BLS1s   & {\bf -0.47}  &   0.25    &  0.13       &    0.22    &    0.09    &    -0.10    &  -0.03 &  -0.15       & 0.02     \\
(3) R4570          &{\bf -0.70}&{\bf -0.38}& all        &{\bf -0.38}& -0.02       &{\bf 0.50}  &{\bf 0.46}  &{\bf -0.47}  &  0.16  &{\bf 0.71}    &{\bf -0.47}\\
                  &{\bf -0.45}&{\bf -0.35}& NLS1s      &    -0.13  &  0.15       &    0.18    &{\bf 0.38}  &{\bf -0.38}  &  0.23  &{\bf  0.53}   &   0.01    \\
                  &{\bf -0.50}&{\bf -0.47}& BLS1s      &    -0.23  & -0.21       &    0.20    &     0.17   &     -0.12   &  0.09  &    0.41      & -0.27     \\
(4) FWHM$_{\rm [SII]}$ &{\bf 0.38}&{\bf 0.29}&{\bf -0.38} &   all     &{\bf 0.41}   &-0.25       &{\bf -0.41} &{\bf 0.41}   &  0.12  &{\bf -0.32}   &{\bf 0.37}\\
                    &  0.02    &   0.30   &   -0.13    &   NLS1s   &   0.28      &   -0.01    &{\bf -0.39} &{\bf  0.40}  & -0.03  &  0.01        &   -0.05   \\
                    &  0.15    &   0.25   &   -0.23    &   BLS1s   &{\bf 0.67}   & -0.03      &   -0.12    &      0.06   &  0.41  &   0.05       & {\bf 0.45} \\
(5) FWHM$_{\rm [OIII]_c}$ & 0.15  &  0.10    & -0.02      &{\bf 0.41} &  all        &{\bf 0.31}  &  -0.03     & 0.01        &{\bf 0.29}& -0.01      &{\bf 0.27}  \\
                       & 0.12  &   0.04   &    0.15    &     0.28  &  NLS1s      &{\bf 0.61}  &  0.15      &    -0.15    &{\bf 0.37}&  0.17      &    0.34    \\
                       & 0.24  &   0.13   &  -0.21     & {\bf 0.67}& BLS1s       &  0.17      &-0.22       &     0.18    &   0.20   & -0.13      &  0.27       \\
(6) $v_{\rm_{[OIII]_c}}$&{\bf -0.52}& -0.04   &{\bf 0.50}  &-0.25      &{\bf 0.31}   &    all     & 0.24       &{\bf -0.27}  &   0.09   &{\bf 0.56}  &{\bf -0.34}\\
                    & -0.15     &  -0.08  &     0.18   &    -0.01  &{\bf 0.61}   &    NLS1s   & 0.14       &    -0.14    &   0.24   &   0.34     &    0.16    \\
                    & -0.07     &  0.22   &   0.20     &   -0.03   & 0.17        &    BLS1s   & 0.04       &    -0.06    &  -0.07   &  0.04      & -0.16       \\
(7) R$_{\rm [SII]}$ &{\bf -0.43} & -0.07   &{\bf 0.46}  &{\bf -0.41}& -0.03       &  0.24      &   all      &{\bf -1.00}  &0.03      &{\bf 0.38}  &{\bf -0.35} \\
                   & -0.11      & -0.09   &{\bf 0.38}  &{\bf -0.39}&  0.15       &    0.14    &  NLS1s     &{\bf -1.00}  &   -0.01  &   0.00     &   -0.02    \\
                   &-0.37       &  0.09   &   0.17     &    -0.12  & -0.22       &    0.04    &  BLS1s     &{\bf -1.00}  &   0.13   &  0.38      & -0.17       \\
(8) $n_{\rm e}$      &{\bf 0.46} &  0.07   &{\bf -0.47} &{\bf 0.41} &  0.01       &{\bf -0.27} &{\bf -1.00} & all         & -0.06    &{\bf -0.42} &{\bf 0.34}\\
                    &  0.11     &   0.09  &{\bf -0.38} &{\bf  0.40}&  -0.15      &  -0.14     &{\bf -1.00} & NLS1s       &   0.01   &  -0.003    &    0.02    \\
                    & 0.38      &  -0.10  &   -0.12    &      0.06 &   0.18      &   -0.06    &{\bf -1.00} & BLS1s       &   -0.23  &{\bf -0.45} & 0.10       \\
(9) $\lambda$L$_{5100}$ &0.04& -0.10   & 0.16       &  0.12     &{\bf 0.29}   &   0.09     &  0.03      &-0.06        &   all    &{\bf 0.34}  &{\bf 0.59}\\
                           &0.22& -0.16   &  0.23      &   -0.03   &{\bf 0.37}   &   0.24     &  -0.01     &    0.01     &    NLS1s &{\bf 0.54}  & {\bf 0.90} \\
                          &-0.13& -0.03   &  0.09      &     0.41  &   0.20      &   -0.07    &  0.13      &   -0.23     &    BLS1s &{\bf  0.57} & {\bf 0.74}  \\
(10) $L/L_{\rm Edd}$ &{\bf -0.91}& -0.14   &{\bf 0.71}  &{\bf -0.32}& -0.01       &{\bf 0.56}  &{\bf 0.38}  &{\bf -0.42}  &{\bf 0.34}& all        &{\bf-0.50}\\
                    &{\bf -0.67}&  0.01   &{\bf 0.53}  &    0.01   &   0.17      &    0.34    &    0.00    &    -0.003   &{\bf 0.54}& NLS1s      &    0.15  \\
                    &{\bf -0.85}& -0.15   &  0.41      &    0.05   &  -0.13      &    0.04    &    0.38    &{\bf -0.45}  &{\bf 0.57}& BLS1s      &   -0.06  \\
(11) $M_{\rm BH}$     &{\bf 0.79}&  0.01   &{\bf -0.47} &{\bf 0.37} &{\bf 0.27}   &{\bf-0.34}  &{\bf -0.35} &{\bf 0.34}   &{\bf 0.59}&{\bf -0.50} &all   \\
                     &{\bf 0.59}& -0.20   &   0.01     &   -0.05   &   0.34      &    0.16    &   -0.02    &     0.02    &{\bf 0.90}&   0.15     &NLS1s  \\   
                     &{\bf 0.51}&  0.02   &  -0.27     &{\bf 0.45} &   0.27      &   -0.16    &   -0.17    &     0.10    &{\bf 0.74}&  -0.06     &BLS1s \\
\enddata
\tablenotetext{a}
{All correlations with Spearman rank probabilities $P_{\rm s} < 0.01$
are marked in boldface.}
\label{table3}
\end{deluxetable}

\clearpage
\begin{deluxetable}{lrrrrrrr}
\tablecaption{Results of the PCA. The relative significances of the 
eigenvectors are listed, 
as well as the projections of the original parameters.}
\tablewidth{0pt}
\tablehead{
\colhead{} & \colhead{EV1} & \colhead{EV2} &\colhead{EV3} &\colhead{EV4} &\colhead{EV5} &\colhead{EV6} &\colhead{EV7}
}
\startdata
Eigenvalue          & 2.6592 & 1.2295 & 1.0299 & 0.6432 & 0.6050 & 0.4366 & 0.3966 \\
Percentage variance & 37.99  & 17.56  & 14.71  & 9.19   & 8.64   & 6.24   & 5.67   \\
Cumulative          & 37.99  & 55.55  & 70.26  & 79.45  & 88.09  & 94.33  & 100.00 \\
\\
$M_{\rm i}$            &   -0.103 &-0.885 & 0.183 &-0.134 &-0.295 & 0.259 &-0.020  \\
FWHM$_{\rm H\beta_{\rm b}}$ &    0.758 &-0.015 &-0.212 &-0.216 & 0.403 & 0.367 & 0.189  \\
R5007                  &    0.441 & 0.004 & 0.841 & 0.038 & 0.059 &-0.132 & 0.276  \\
R4570                  &   -0.759 & 0.331 &-0.062 & 0.173 &-0.229 & 0.277 & 0.390  \\
FWHM$_{\rm [SII]}$     &    0.667 & 0.451  &0.238 & 0.175 &-0.304 & 0.308 &-0.278  \\
$v_{\rm_{[OIII]_c}}$   &   -0.606 & 0.347  &0.298 &-0.634 & 0.025 & 0.080 &-0.122  \\
R$_{\rm [SII]}$        &   -0.700 &-0.113 & 0.309 & 0.339 & 0.454 & 0.200 &-0.200  \\
\enddata
\label{table4}
\end{deluxetable}

\begin{deluxetable}{lrr}
\tablecaption{Spearman rank order correlation for the first two eigenvectors
with $n_{\rm e}$, $\lambda L_{5100}$, $M_{\rm BH}$ and $L/L_{\rm Edd}$
\tablenotemark{a}} 
\tablewidth{0pt}
\tablehead{
\colhead{} & \colhead{EV1} & \colhead{EV2}}
\startdata
$n_{\rm e}$        &  {\bf 0.71}  &   0.05      \\
$\lambda L_{5100}$ &      -0.01   & {\bf 0.86}  \\
$M_{\rm BH}$       &  {\bf 0.60}  & {\bf 0.48}  \\
$L/L_{\rm Edd}$    &  {\bf -0.71} & {\bf 0.36}  \\
\enddata
\tablenotetext{a}
{All correlations with Spearman rank probabilities $P_{\rm s} < 0.01$
are marked in boldface.}
\label{table5}
\end{deluxetable}

\clearpage
\onecolumn
\begin{figure*}
\plotone{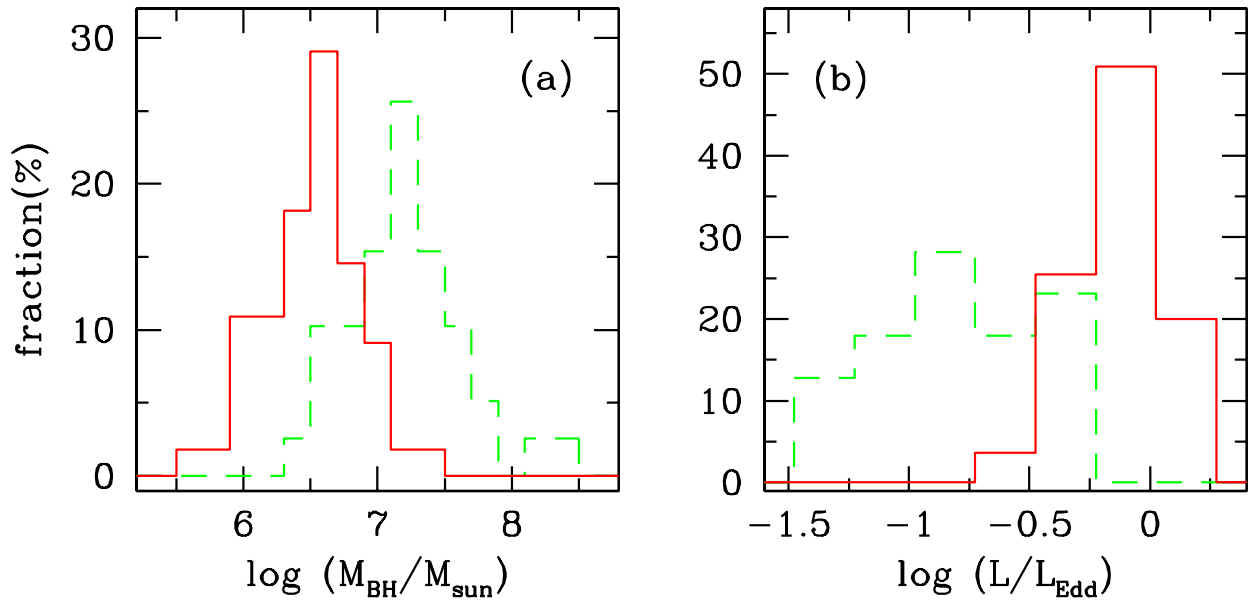}
\caption{ 
Distribution of black hole masses and Eddington ratios
of the galaxies of our sample.
NLS1 galaxies are shown as solid line, 
BLS1 galaxies as dashed line.}
\label{fig1}
\end{figure*}

\clearpage
\begin{figure*}
\plotone{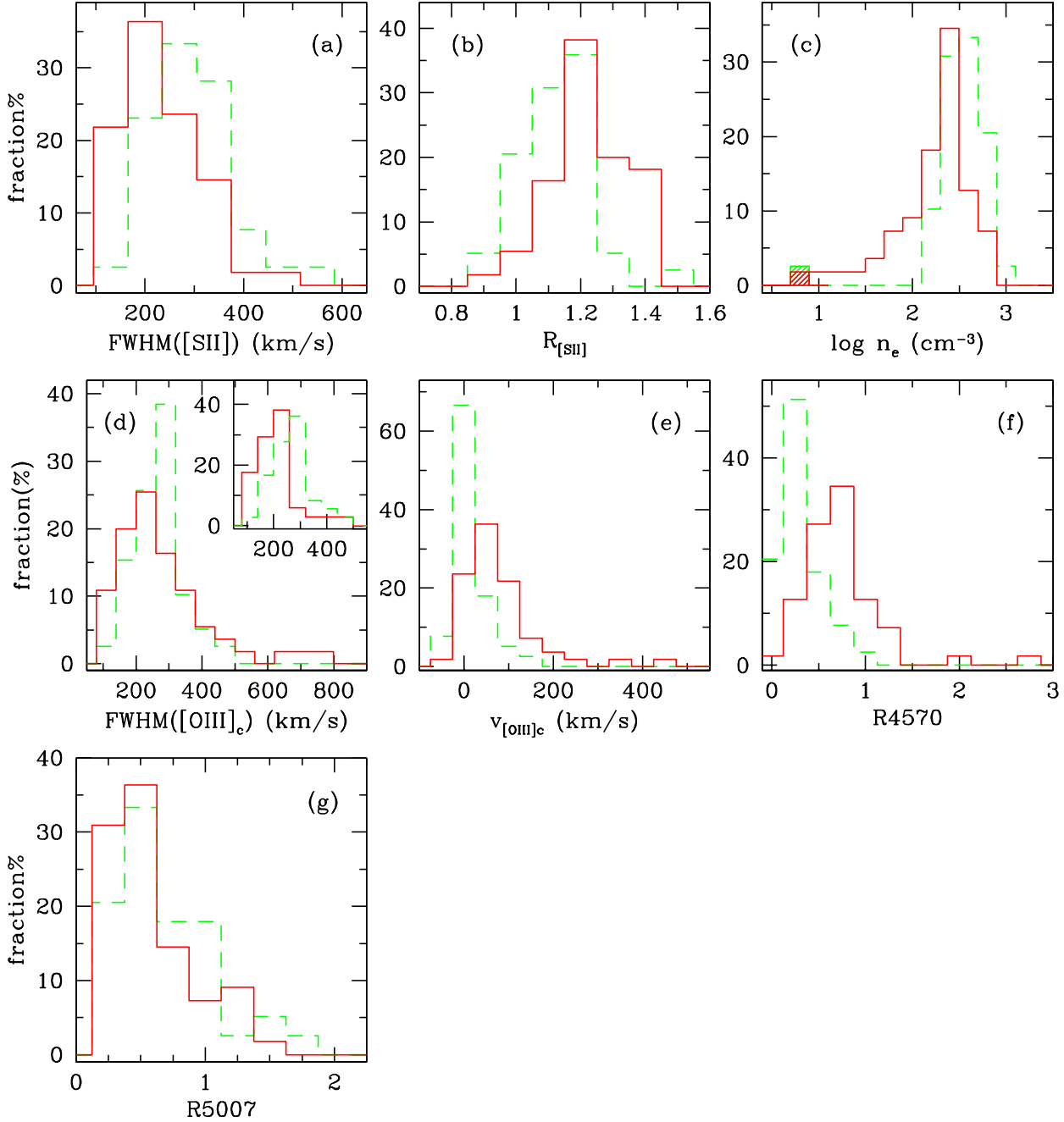}
\caption{Histograms showing the distribution of the sample properties of 
NLS1 and BLS1 galaxies. 
NLS1 properties are shown as solid line, BLS1 properties as dashed line.
Panels from left to right and top to bottom:
{\em (a)} FWHM(\sii) in \kms; 
{\em (b)} intensity ratio of [SII]$\lambda6716/\lambda6731$;
{\em (c)} logarithm of NLR electron density in cm$^{-3}$.
     The shaded histogram indicates the two galaxies with 
     \n $<$ 10\,\cm. 
{\em (d)} FWHM(\oiii$_{\rm c}$) in \kms. 
     The inset shows the distribution of \oiii$_{\rm c}$ 
     velocity (blueshift) after excluding objects in the 
     highest velocity bin;
{\em (e)} \oiii$_{\rm c}$ velocity (blueshift) w.r.t. \sii\ in km\,s$^{-1}$;
{\em (f)} ratio of \feii\,4570 over total H$\beta$ emission;
{\em (g)} ratio of total \oiii\ over total H$\beta$ emission.
}
\label{fig2}
\end{figure*}

\clearpage
\begin{figure*}
\plotone{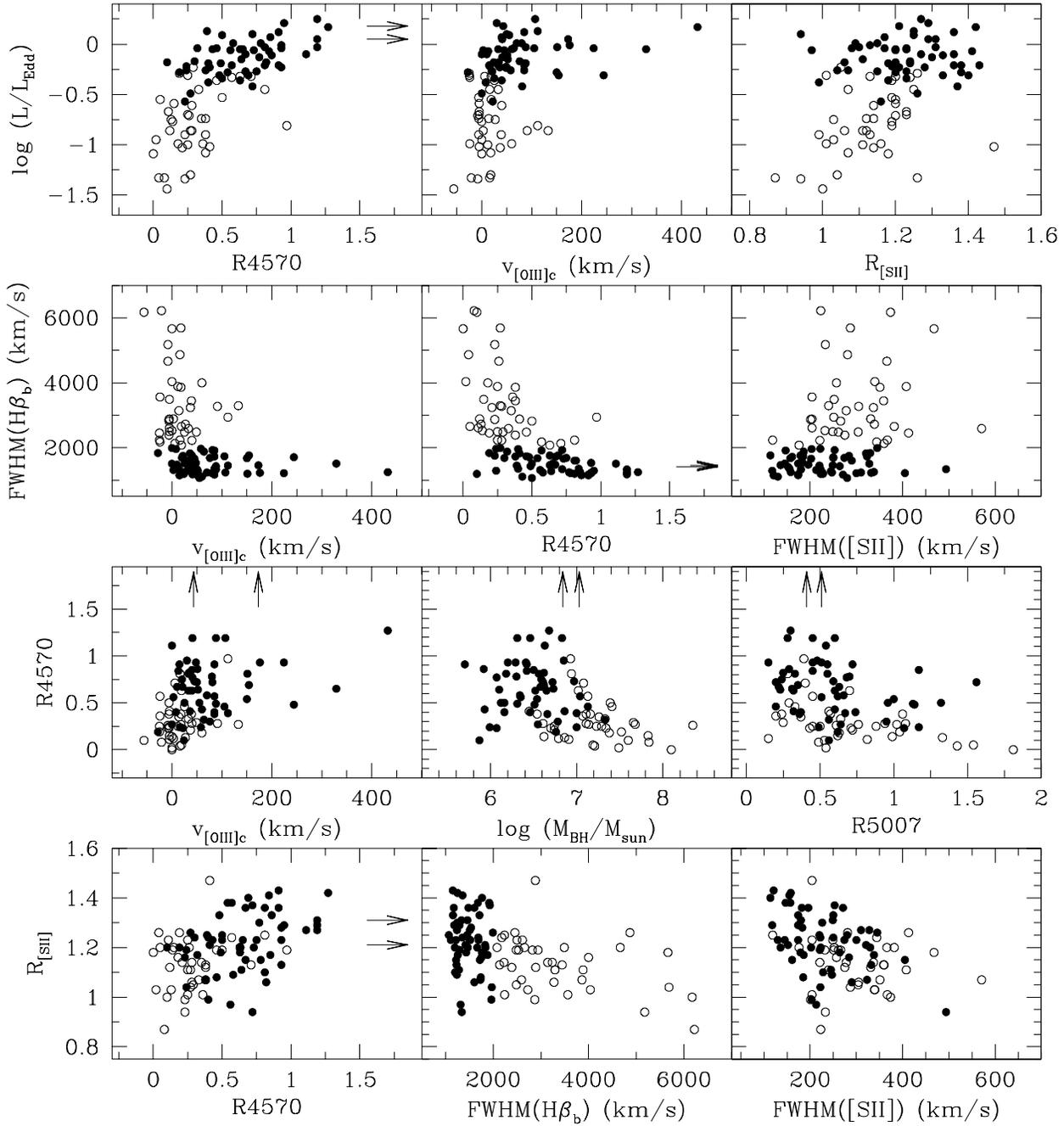}
\caption{Strongest correlations among emission-line and continuum properties. 
NLS1 galaxies are coded as filled circles and BLS1 galaxies as open circles.
The objects that are off the plots are indicated by arrows.}
\label{fig3}
\end{figure*}

\clearpage
\begin{figure*}
\plotone{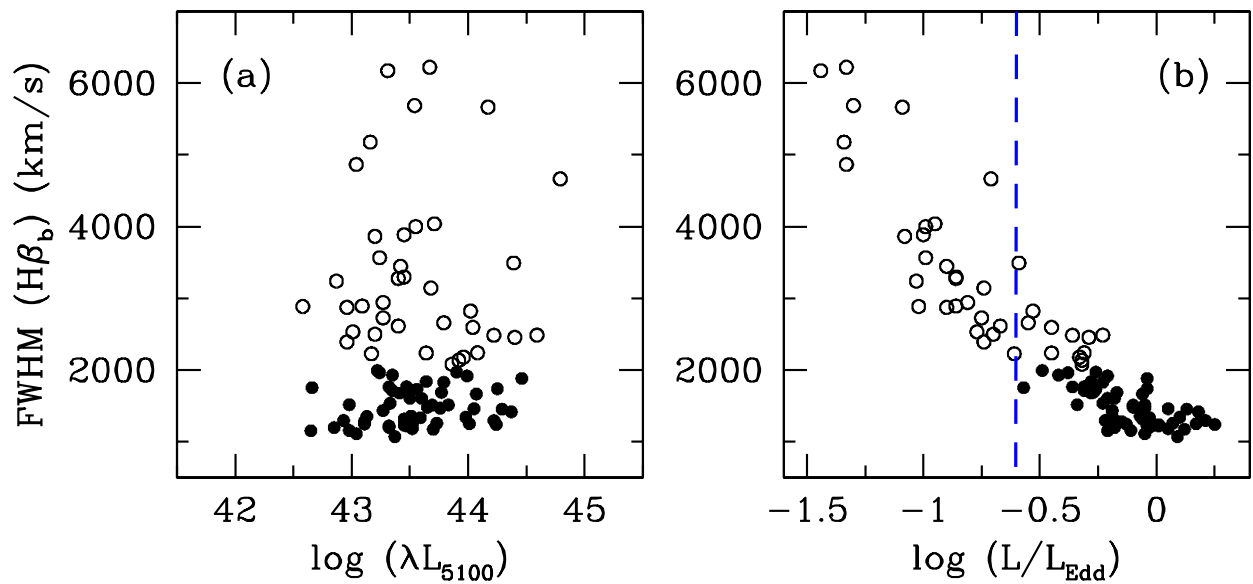}
\caption{
FWHM of the broad component of H$\beta$
versus $\lambda L_{5100}$ and $L/L_{\rm Edd}$.
Symbols are as in Fig.\,3. The dashed line
in the right panel corresponds to $L/L_{\rm Edd} = 0.25$.
}
\label{fig4}
\end{figure*}

\clearpage
\begin{figure*}
\plotone{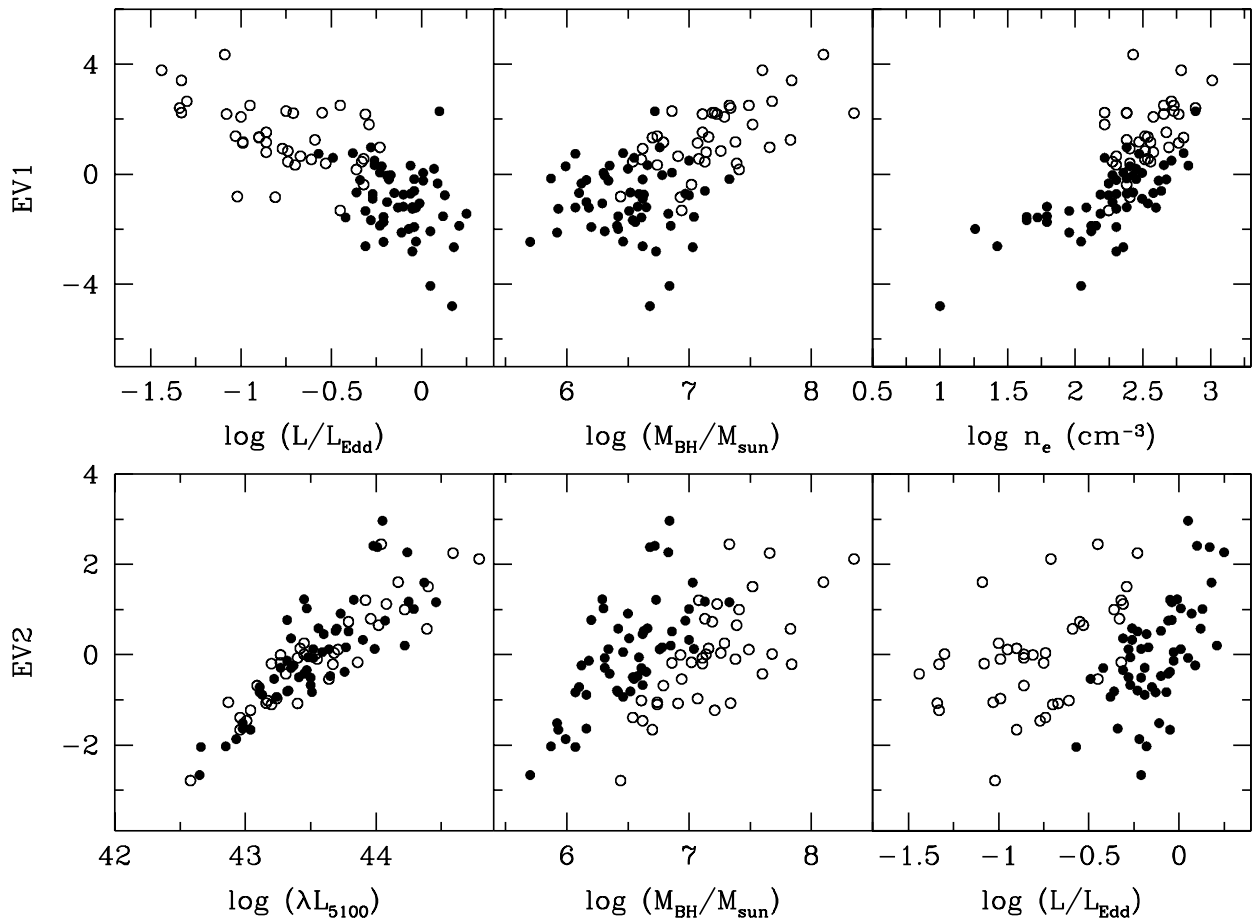}
\caption{Correlations of the first two eigenvectors with 
$L/L_{\rm Edd}$, $M_{\rm BH}$, $n_{\rm e}$, and $\lambda L_{5100}$.
Symbols are as in Fig.\,3. The eigenvectors are plotted in arbitrary
units. 
}
\label{fig5}
\end{figure*}

\clearpage
\begin{figure*}
\plotone{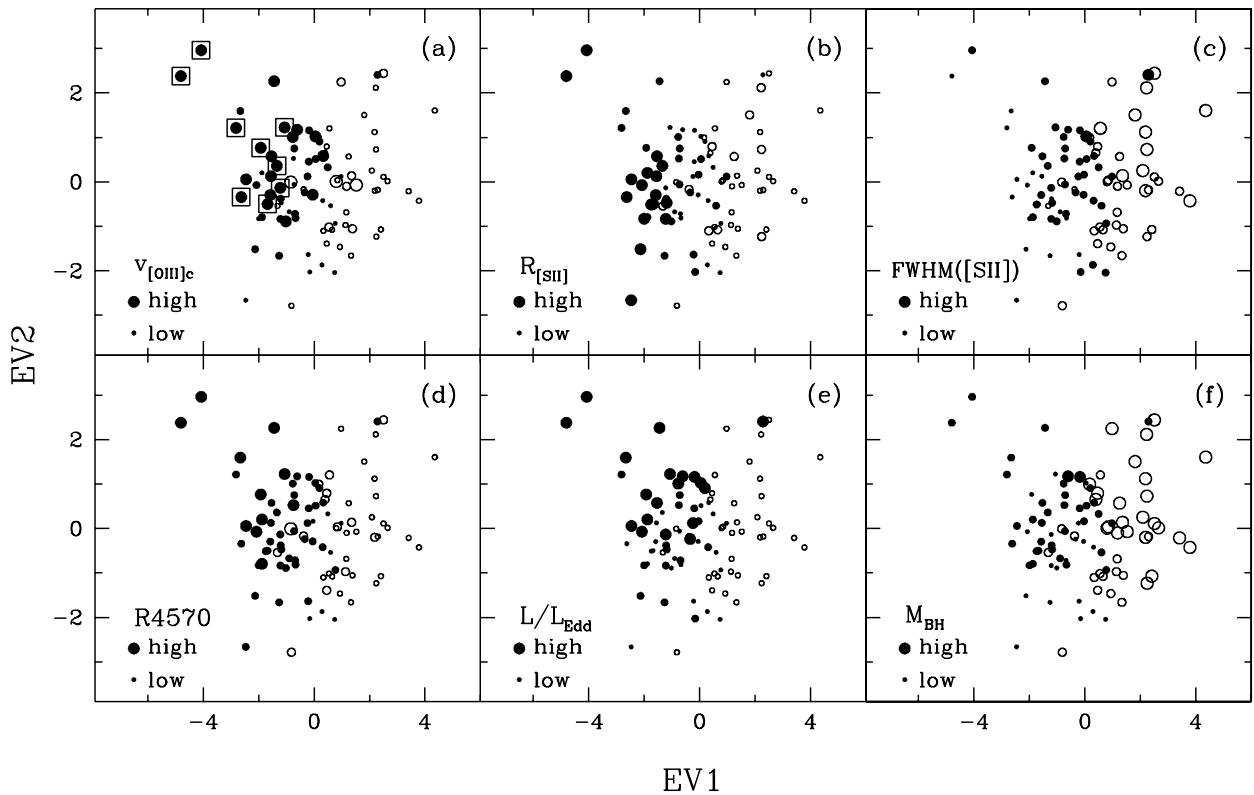}
\caption{Influence of different parameters on the distribution of 
NLS1 galaxies (filled circles) and BLS1 galaxies (open circles) 
with respect to the first two eigenvectors. Each parameter is
divided into three bins and coded by circle size: 
large circles, high value; medium circles, intermediate value; 
small circles, low value. 
Left to right and top to bottom: coding according to 
\oiii$_{\rm c}$ blueshift (blue outliers in \oiii$_{\rm c}$ 
are marked with an extra open square.), R$_{\rm [SII]}$, FWHM([SII]), 
R4570, $L/L_{\rm Edd}$, $M_{\rm BH}$. 
}
\label{fig6}
\end{figure*}

\clearpage
\begin{figure}
\plotone{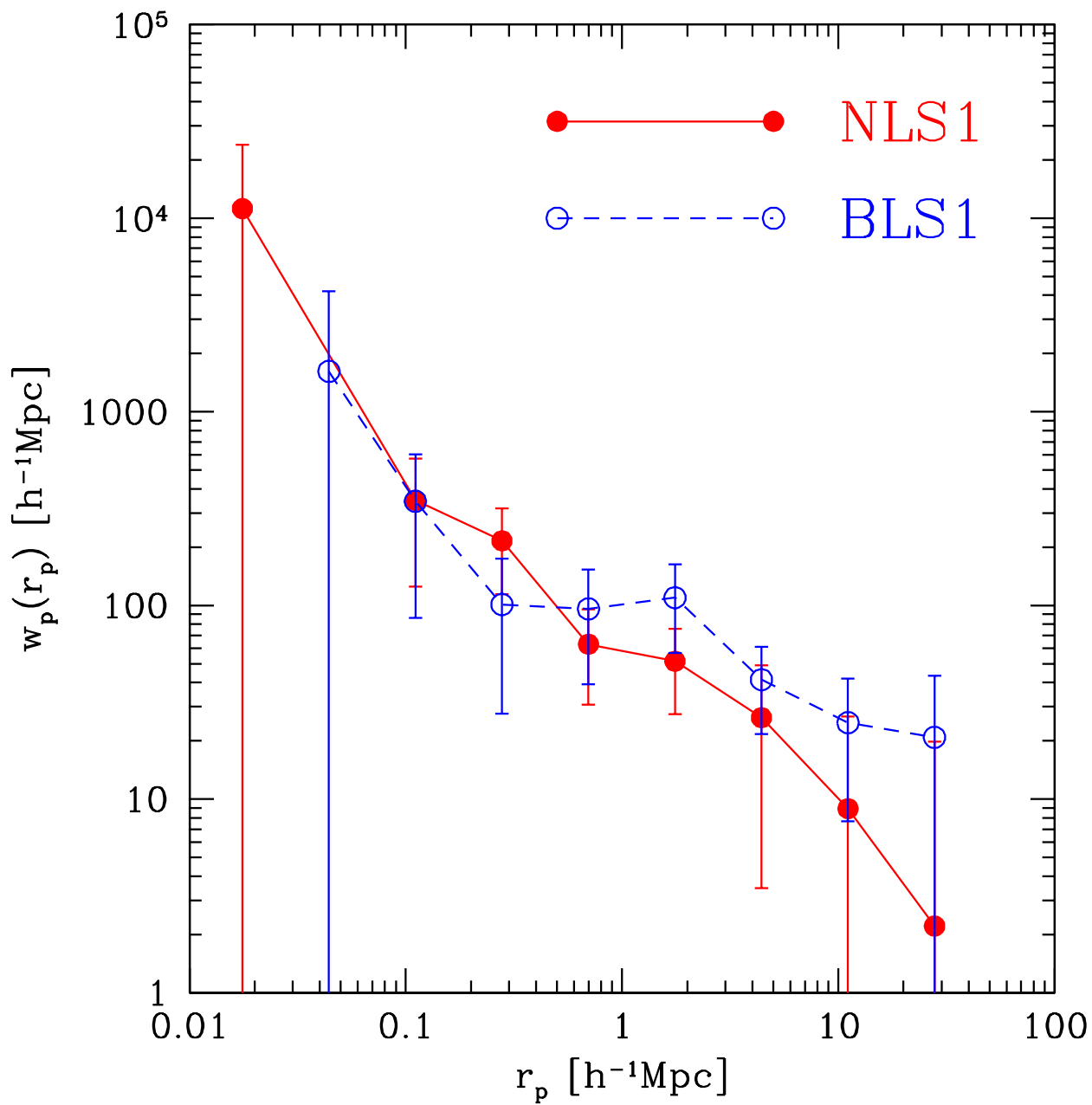}
 \caption{Projected cross-correlation function $w_p(r_p)$
for NLS1 galaxies (filled circles connected by the solid line) 
and BLS1 galaxies (open circles connected by the dashed line)
in our samples, with respect to a reference sample of about 
half a million galaxies in the SDSS/DR7. Errors in the $w_p(r_p)$
measurements are estimated from the Bootstrap resampling
technique.}
 \label{fig:wrp}
\end{figure}


\begin{thebibliography}{}
\bibitem[Abazajian et al.(2003)]{dr1} Abazajian, K., 
et al. 2003, \aj, 126, 2081
\bibitem[Abazajian et al.(2005)]{dr3} Abazajian, K., 
et al. 2005, \aj, 129, 1755
\bibitem[Abazajian et al.(2009)]{dr7} Abazajian, K., et al. 2009,
\apjs, 182, 543
\bibitem[Abdo et al. (2009)]{Abdo} Abdo, A. A., et al. 2009, \apj, 707, L 142
\bibitem[Adelman-McCarthy et al.(2008)]{dr6} Adelman-McCarthy, J. K., 
  Ag{\"u}eros, M. A., Allam, S. S., et al. 2008, \apjs, 175, 297 
\bibitem[Bentz et al.(2009)]{bentz2009} Bentz, M. C., Peterson, B. M., 
Netzer, H., Pogge, R. W., \& Vestergaard, M. 2009, \apj, 697, 160
\bibitem[Boroson \& Green(1992)]{bg92} Boroson, T. A., \& Green, R. F. 1992, 
  \apjs, 80, 109
\bibitem[Boroson(2002)]{boroson02} Boroson, T. A. 2002, \apj, 565, 78
\bibitem[Boroson(2003)]{boroson03} Boroson, T. A. 2003, \apj, 585, 647
\bibitem[Boroson 2004]{boroson04} Boroson, T. A. 2004, ASP Conf. Ser., 311, 3
\bibitem[Brandt(1999)]{brandt99} Brandt, W. N. 1999, ASP Conf. Ser., 161, 166
\bibitem[Brandt \& Gallagher(2000)]{brandt00} Brandt, W. N., \& 
Gallagher, S. C. 2000, NewAR, 44, 461
\bibitem[Bruzual \& Charlot(2003)]{bc03} Bruzual, D., \& Charlot, S. 2003,
\mnras, 344, 1000
\bibitem[Centrella et al.(2010)]{centrella10} Centrella, J., 
Baker, J. G., Kelly, B. J., \& van Meter, J. R. 2010, Rev. Mod. Phys.,
82, 3069
\bibitem[Crenshaw et al.(2003)]{crenshaw03} Crenshaw, D. M., 
Kraemer, S. B., \&  Gabel, J. R. 2003, \aj, 126, 1690	
\bibitem[Deo et al.(2006)]{deo06} Deo, R. P., Crenshaw, D. M., 
\& Kraemer, S. B. 2006, \aj, 132, 321
\bibitem[Dietrich et al.(2003)]{dietrich03} Dietrich, M., Hamann, F., 
Appenzellar, I., \& Vestergaard, M. 2003, \apj, 596, 817
\bibitem[Dietrich et al.(2005)]{dietrich05} Dietrich, M., Grenshaw, D. M., \& Kraemer, S. B. 2005, \apj, 623, 700
\bibitem[di Matteo, Springel, \& Hernquist(2005)]{dimatteo05} 
di Matteo, T., Springel, V., \& Hernquist, L. 2005, \nat, 433, 604
\bibitem[Duda et al. (2001)]{duda01} Duda, R. O., Hard, P. E., \& Stork,  
D. G., 2001, ``Pattern Clasification'', Wiley \& Sons
\bibitem[Foschini(2011)]{foschini11} 
Foschini, L. 2011, arXiv:1105.0772, Proceedings of the conference
``Narrow-Line Seyfert 1 Galaxies and their place in the Universe'', 
published online at http://pos.sissa.it/cgi-bin/reader/conf.cgi?confid=126
\bibitem[Francis \& Wills(1999)]{francis99} Francis, P. J., \& 
Wills, B. J. 1999, ASP Conf. Series, 162, 363 
\bibitem[Goodrich(1989)]{good89} Goodrich, R. W. 1989, \apj, 340, 190
\bibitem[Grupe et al.(1999)]{gru99} Grupe, D., Beuermann, K., Mannheim, K., 
  \& Thomas, H.-C. 1999, \aap, 350, 31
\bibitem[Grupe(2004)]{gru04} Grupe, D. 2004, \aj, 127, 1799
\bibitem[Grupe et al.(2010)]{gru10} Grupe, D., Komossa, S., 
Leighly, K. M., \& Page, K. L. 2010, \apjs, 187, 64
\bibitem[Hogg et al.(2001)]{Hogg01} Hogg, D. W., Finkbeiner, D. P., 
  Schlegel, D. J., \& Gunn, J. E. 2001, \aj, 122, 2129
\bibitem[Kaiser et al.(2000)]{kaiser00} 
Kaiser, M. E., et al., 2000, \apj, 528, 260
\bibitem[Kaspi et al.(2000)]{kaspi00} Kaspi, S., Smith, P. S., Netzer, H., 
  et al. 2000, \apj, 533, 631
\bibitem[Kaspi et al.(2005)]{kaspi05} Kaspi, S., Maoz, D., Netzer, H., 
  et al. 2005, \apj, 629, 61
\bibitem[Komossa et al.(2006)]{stefanie06} Komossa, S., Voges, W., Xu, D., 
 Mathur, S., Adorf, H.-M., Lemson, G., Duschl, W., \& Grupe, D. 2006, \aj, 
 132, 531 
\bibitem[Komosssa \& Xu 2007]{kx07} Komossa, S., \& Xu, D. 2007, \apj, 667, L33 (KX07)
\bibitem[Komossa et al.(2008)]{komossa08a} Komossa, S., Xu, D., Zhou, H.,
    Storchi-Bergmann, T., \& Binette, L. 2008, \apj, 680, 926 (K08) 
\bibitem[Komosssa 2008]{komossa08b} Komossa, S. 2008, RevMexAA 
    (Serie de Conferencias), 32, 86 
\bibitem[Kormendy \& Kennicutt(2004)]{kormendy04} 
Kormendy, J., \& Kennicutt, R. C., Jr. 2004, \araa, 42, 603	
\bibitem[Kraemer, Schmitt, \& Crenshaw(2008)]{kraemer08} 
Kraemer, S. B., Schmitt, H. R., \& Crenshaw, D.M. 2008, \apj, 679, 1128
\bibitem[Kriss(1994)]{kriss94} Kriss, G. A. 1994, ASP Conf. Ser. 61, 437
\bibitem[Krongold et al.(2001)]{kron01} Krongold, Y., Dultzin-Hacyan, D., 
\& Marziani, P. 2001, \aj, 121, 702 
\bibitem[Kuraszkiewicz et al(2000)]{kur00} Kuraszkiewicz, J., 
et al. 2009, \apj, 692, 1180
\bibitem[Laor et al.(1997)]{lao97} Laor, A., Fiore, F., Martin, E., et al. 
  1997, \apj, 477, 93
\bibitem[Laor 2000]{laor2000} Laor, A. 2000, NewAR, 44, 503 
\bibitem[Li et al.(2006)]{li06} Li, C., et al. 2006, \mnras, 373, 457
\bibitem[Li et al.(2008)]{li08} Li, C., et al. 2008 \mnras, 385, 1903
\bibitem[Lu et al.(2006)]{lu06} Lu, H., Zhou, H., Wang, T., Dong, X., \& 
  Li, C. 2006, \aj, 131, 790
\bibitem[Maciejewski et al.(2002)]{maciejewski02} 
Maciejewski, W., Teuben, P. J., Sparke, L. S., \& Stone, J. M. 2002,
\mnras, 329, 502 
\bibitem[Maciejewski(2004a)]{maciejewski04a} Maciejewski, W. 2004a,
\mnras, 354, 883
\bibitem[Maciejewski(2004b)]{maciejewski04b} Maciejewski, W. 2004b,
\mnras, 354, 892
\bibitem[Mao, Wang, \& Wei{2009}]{mao09} 
Mao, Y.-F., Wang, J., \& Wei, J. Y. 2009, RAA, 9, 529
\bibitem[Mathur, Kuraszkiewicz \& Czerny(2001)]{mat01} Mathur, S., 
 Kuraszkiewicz, J., \& Czerny, B. 2001, NewA, 6, 321
\bibitem[Mathur et al.(2011)]{mathur11} Mathur, S., et al., 2011, 
arXiv:1102.0537
\bibitem[Marziani et al.(2001)]{mar01} Marziani, P., Sulentic, J. W., 
  Zwitter, T., Dultzin-Hacyan, D., \& Calvai, M. 2001, \apj, 558, 553
\bibitem[Marziani et al.(2003)]{mar03} Marziani, P., Zamanov, R. K., 
   Sulentic, J. W., \& Calvani, M. 2003, \mnras, 345, 1133
\bibitem[Marziani et al.(2010)]{mar10} Marziani, P., et al. 2010,
\mnras, 409, 1033
\bibitem[Morganti et al.(2010)]{Morganti10} 
Morganti, R., Holt, J., Tadhunter, C., \& Oosterloo, T. 2010, 
IAU Symp. 267, 429
\bibitem[Mullaney \& Ward(2008)]{mull08} Mullaney, J. R., \& Ward, M. J.
    2008, \mnras, 385, 53
\bibitem[Netzer \& Trakhtenbrot(2007)]{netzer07} Netzer, H., \& Trakhtenbrot, B. 2007, \apj, 654, 754
\bibitem[Ohta et al.(2007)]{ohta07} Ohta, K., Aoki, K., Kawaguchi, T.,
  \& Kiuchi G. 2007, \apjs, 169, 1
\bibitem[Osterbrock \& Pogge(1985)]{ost85} Osterbrock, D. E., \& Pogge, R. W. 
  1985, \apj, 297, 166
\bibitem[Patsis \& Athanassoula(2000)]{patsis00}Patsis, P. A., 
\& Athanassoula, E. 2000, \aap, 358, 45
\bibitem[Peason (1901)]{peason01} Peason, K. 1901, Philosophical  
Magazine 2, 559
\bibitem[Peterson et al.(2000)]{peterson00} Peterson, B. M., et al. 2000, 
  \apj, 542, 161
\bibitem[Peterson et al.(2004)]{peterson04} Peterson, B. M., et al. 2004,
  \apj, 613, 682
\bibitem[Proga., Ostriker, \& Kurosawa(2010)]{Proga10}
Proga., D., Ostriker, J., \& Kurosawa, R. 2010, \apj, 676, 101
\bibitem[Riffel et al.(2008)]{Riffel08}
Riffel, R., et al. 2008, MNRAS, 385, 1129
\bibitem[Rodriguez-Ardila et al.(2000)]{rod00} Rodriguez-Ardila, A., 
  Binette, L., Pastoriza, M. G., \& Donzelli, C. J. 2000, \apj, 538, 581
\bibitem[Ryan et al.(2007)]{ryan07} Ryan, C. J., De Robertis, M. M., 
Virani, S., Laor, A., \& Dawson, P. C. 2007, \apj, 654, 799
\bibitem[Schiano(1986)]{schiano86} Schiano A. V. R. 1986, ApJ, 302, 81
\bibitem[Shlosman, Begelman \& Frank(1990)]{shl90} Shlosman, I.,
  Begelman, M. C., \& Frank, J. 1990, \nat, 345, 679
\bibitem[Shemmer et al.(2004)]{shemmer00} Shemmer, O., Netzer, H., 
    Mailolino, R., et al. 2004, \apj, 614, 547
\bibitem[Sulentic et al.(2000)]{sulentic00}
 Sulentic, J.W., Zwitter, T., Marziani, P., \& Dultzin-Hacyan, D. 2000, 
  \apj, 536, L5
\bibitem[Sulentic et al.(2002)]{sulentic02} Sulentic, J. W., Marziani, P.,
 Zamanov, R., Bachev, R., Calvani, M., \& Dultzin-Hacyan, D. 2002, 
 \apj, 566, L71 
\bibitem[Sulentic et al. 2007]{sulentic07} Sulentic, J., Bachev, R., 
    Marziani, P., Negrete, C.A., \& Dultzin, D. 2007, \apj, 666, 757
\bibitem[Sulentic et al.(2008)]{sulentic08} Sulentic, J. W., Zamfir, S., 
     Marziani, P, \& Dultzin, D. 2008, RevMexAA (Serie de Conferencias), 
     32, 51   
\bibitem[Vaughan et al.(2001)]{vau01} Vaughan, S., Edelson, R., Warwick, R. S., 
  et al. 2001, \mnras, 327, 673
\bibitem[V\'{e}ron-Cetty, V\'{e}ron \& Gon\c{c}alves(2001)]{veron01} 
  V\'{e}ron-Cetty, M. P., V\'{e}ron, P., \& Gon\c{c}alves, A. C. 2001, 
  \aap, 372, 730 
\bibitem[V\'{e}ron-Cetty \& V\'{e}ron(2003)]{veron00} V\'{e}ron-Cetty, M. P.,
  \& V\'{e}ron, P. 2003, \aap, 412, 399
\bibitem[V\'{e}ron-Cetty, Joly \& V\'{e}ron(2004)]{veron04} V\'{e}ron-Cetty,
 M. P., Joly, M., \& V\'{e}ron, P. 2004, \aap, 417, 515
\bibitem[Vestergaard(2004)]{vestergaard04} Vestergaard, M. 2004, \apj, 601, 676
\bibitem[Wandel et al.(1999)]{wandel99} Wandel, A., Peterson, B. M., \& 
  Malkan, M. A. 1999, \apj, 526, 579  
\bibitem[Wang et al.(2010)]{wang10} 
Wang, J., et al. 2010, ApJ, 719, L208
\bibitem[Wang et al.(2006)]{Wang06} 
Wang, J., Wei, J. Y., \& He, X. T. 2006, \apj, 638, 106
\bibitem[Weymann et al.(1991)]{weymann91} 
Weymann, R. J., Morris, S. L., \& Foltz, C. B. 1991, \apj, 373, 23
\bibitem[Williams, Mathur, \& Pogge(2004)]{wil04}
   Williams, J. M., Mathur, S., \& Pogge, R. W. 2004, \apj, 610, 737
\bibitem[Wu \& Liu(2004)]{wu04} Wu, X. B., \& Liu, F. K. 2004, \apj, 614, 91
\bibitem[Orban de Xivry et al.(2011)]{Xivry11} 
Orban de Xivry, G., et al. 2011, \mnras, 417, 2721
\bibitem[Xu et al.(2007)]{xuetal07} Xu, D., Komossa, S., Zhou, H.,
   Wang, T., \& Wei, J. 2007, \apj, 670, 60 (X07)
\bibitem[Yuan et al.(2008)]{yuan08} 
Yuan, W., et al. 2008, \apj, 685, 801
\bibitem[Zamfir, Sulentic, \& Marziani(2008)]{zamfir08}
Zamfir, S., Sulentic, J. W., \& Marziani, P. 2008, MNRAS, 387, 856
\bibitem[Zhang et al.(2011)]{zhang11} Zhang, K., Dong, X., Wang, T.,
\& Gaskell, C. M. 2011, \apj, 737, 71 
\bibitem[Zhou et al.(2003)]{zhou03} Zhou, H.,  Wang, T., Dong, X., 
Zhou, Y., \& Li, C.  2003, \apj, 584, 147
\bibitem[Zhou et al.(2006)]{zhou06} Zhou, H., Wang, T., Yuan, W., Lu, H.,
  Dong, X., Wang, J. \& Lu, Y. 2006, \apjs, 166, 128 
\end{thebibliography}
\end{document}